\documentclass[a4paper,10pt]{article}
\usepackage[utf8]{inputenc}
\usepackage{hyperref}
\usepackage{amsfonts}
\usepackage{amsmath}
\usepackage{amssymb}
\usepackage{indentfirst}
\usepackage{graphicx}
\usepackage{cite}
\usepackage{color}
\usepackage{xcolor}

\usepackage{breqn}

\usepackage{setspace}%

\hypersetup{
colorlinks=true,
citecolor=black,
linkcolor=black,
filecolor=black,
urlcolor=black,
}
\numberwithin{equation}{section}
\allowdisplaybreaks
\makeatletter

\begin{document}

\begin{titlepage}
\rightline{MPP-2021-124} 

\vskip 2cm

\begin{center}
\LARGE{\textbf{On the }$L_\infty$\textbf{ formulation of Chern--Simons theories }}
\par\end{center}{\LARGE \par}

\begin{center}
	\vspace{1cm}
	\textbf{S. Salgado}
	\small
	\\[6mm]
	\textit{Max-Planck-Institut f\"{u}r Physik, F\"{o}hringer Ring 6, 80805 Munich, Germany }\\[0.5mm]
	\textit{Ludwig-Maximilians-Universit\"{a}t M\"{u}nchen, Theresienstra\ss e 37, 80333 Munich, Germany}
\textit{ }
	 \\[3mm]
	\footnotesize
	\texttt{E-mail: salgado@mppmu.mpg.de}
	 \\[5mm]
	\today
 
	\par\end{center}
\vskip 20pt
\centerline{{\bf Abstract}}
\medskip
\noindent $L_{\infty}$ algebras have been recently studied as algebraic frameworks in the
formulation of gauge theories in which the gauge symmetries and the dynamics
of the interacting theories are contained in a set of products acting on a
graded vector space. On the other hand, FDAs are differential algebras that
generalize Lie algebras by including higher-degree differential forms on their
differential equations. In this article, we review the dual relation between
FDAs and $L_{\infty}$ algebras. We study the formulation of standard
Chern--Simons theories in terms of $L_{\infty}$ algebras and extend the
results to FDA-based gauge theories. We focus on two cases, namely a flat (or zero-curvature)
theory and a generalized Chern--Simons theory, both including high-degree
differential forms as fundamental fields.

\end{titlepage}\newpage {} 

\noindent\rule{162mm}{0.4pt}
\tableofcontents
\noindent\rule{162mm}{0.4pt}

\section{Introduction}

Using algebras is a way to mathematically express the symmetries of a
physical system. When the physical system is described by physical theory,
we have a set of symmetries that belong to the theory itself and these are
usually such that they can be expressed as Lie algebras. One well-known
exception to this is general relativity, which is globally invariant under
diffeomorphisms. Such diffeomorphisms leave general relativity invariant but they do not constitute a Lie group. The gauge symmetry is
usually encoded in the gauge algebra, a Lie algebra in most cases, defining
a redundance on the formulation, in general necessary for a covariant
description. The information about physical interaction is contained in the
equations of motion. In contrast, a formulation of gauge theories in terms of a single
mathematical structure was studied on Ref. \cite{Zwie,boots}. This means
that the complete information of the theory, namely, its
definition of gauge transformations (and a definition of covariant
derivative through it), the gauge algebra and the dynamics is included into an algebra that
satisfies an enlarged version of the Jacobi identity, being the gauge
algebra a certain subalgebra.

$L_{\infty}$ algebras are generalizations of Lie algebras in which the
antisymmetric bilinear Lie product is replaced by a set of multilinear
products satisfying a set of identities that generalize the standard Jacobi
identity. In the simplest case, an $L_{\infty}$ algebra is reduced to a Lie
algebra. However, the presence of higher products allows the bilinear
product not to satisfy the standard Jacobi identity, and therefore, a
non-associative realization of the bilinear product is possible \cite%
{lada,lada2}. In most cases, when using an $L_{\infty}$ algebra to describe
a gauge theory, the gauge subalgebra $L_{\infty}^{\mathrm{gauge}}$ is a Lie
algebra, as expected, being the dynamics of the theory encoded into the
remaining subspace. However, this is not always the case. An example can be
found in higher gauge theories \cite{baez}. Those are generalizations of the
standard gauge theories that involve higher-degree tensors as fundamental
fields, describing the dynamics of extended objects, such as string and
branes, in a similar manner in which a standard gauge theory describes the
dynamics of point particles. As a consequence of including higher-degree
tensors as gauge fields, a higher gauge theory can show an enlarged
structure even in the gauge subalgebra when it is formulated as an $%
L_{\infty}$ algebra. The simplest case of a higher gauge theory, known as $%
p$-form electrodynamics, was introduced in Ref. \cite{teit} as a model in
which the fundamental field is $p$-form evaluated on a Lie algebra. It was
proved that the resulting theory is inconsistent in describing the parallel
transport of extended objects because of its non-invariance under
reparametrizations. That problem is immediately solved by removing the
structure constants of the Lie algebra, making possible a gauge theory with $%
p$-forms for abelian groups.

Free differential algebras (FDAs), also known as Chevalley--Eilenberg
algebras of Lie algebras, are differential algebras whose elements are dual
to the Lie-valuated algebraic elements and whose differential operator
encodes the information of the Lie product. They were first introduced in
the context of supergravity in Ref \cite{dauria1}. As happens with Lie
algebras, FDAs can be gauged by considering non-vanishing field strengths
(or curvatures), allowing the construction of invariant densities and giving
rise to generalized Chern--Simons (CS) forms depending on higher-degree
differential forms as gauge fields.

In standard gauge theory, CS (and transgression) forms are usually chosen as
candidates to Lagrangians, leading to background-free theories depending
only on a one-form gauge field. Examples of these are the well-known 3D CS
theories invariant theories under the Poincare and AdS Lie groups \cite%
{achu,nie,wi1} and their generalizations to supersymmetric and
higher-dimensional cases \cite{cham,cham2,ban,tron2}. Moreover, CS theories that are invariant under FDAs have been studied in Refs. \cite{savv1,savv2,snpb} for
particular cases in which the number of high-order gauge fields is truncated
or in which the structure constants of the gauge algebra are restricted to
particular cases.

As we will specify later, there is a dual relation between $L_{\infty}$
algebras and FDAs, being the first ones represented by a set of $n$-linear
products acting on a graded vector space and satisfying a generalized
version of the Jacobi identity and the latter defined through a set of
generalized Maurer--Cartan equations that extend the standard Maurer--Cartan
equations for Lie algebras. To extend a Lie algebra to an FDA is possible through the inclusion of new
differential equations for higher-degree differential forms that cannot be
trivially split into wedge products of one-forms (this would make the FDA
equivalent to a Lie algebra) due to the presence of non-trivial cocycles,
representatives of the Chevaley--Eilenberg cohomology classes\ of the
original Lie algebra \cite{cohom}.

This work aims to study the relation between FDAs, higher (and standard)
gauge theories and $L_{\infty}$ algebras. The paper is organized as follows:
In Section 2, we briefly review $L_{\infty}$ algebras in the so-called $\ell 
$-picture and the general formulation of gauge theories in terms of $%
L_{\infty}$ algebras \cite{Zwie}. In Section~3, we review FDAs, their gauging, the
construction of transgression and CS forms from Ref. \cite{CSFDA} and the
duality between FDAs and $L_{\infty}$ algebras. In Section 4, we formulate
CS theories for Lie algebras in the $L_{\infty}$ formulation of gauge
theories. In Section 5, we write down the most simple FDA-based theory in
the $L_{\infty}$ formalism, namely, a flat theory whose dynamics is
determined by the zero-curvature conditions. In Section 6, we extend the
results from Section 4 and formulate the generalized CS theories for FDAs as 
$L_{\infty}$ algebras. There are also two appendices with details on the
notation and further calculations.

\section{$L_{\infty}$ algebras}

$L_{\infty}$ algebras provide a framework to formulate gauge theories,
containing in their mathematical structure not only the gauge symmetries but
also the dynamics of the interacting theory, being each one codified in the products of elements from specific
subspaces. This section will shortly review the definition of $L_{\infty}$
algebras in the so-called $\ell$-picture and the procedure to write down the
relevant information of an arbitrary gauge theory in terms of its products 
\cite{Zwie, boots}.

\subsection{Definition}

An $L_{\infty}$ algebra is defined as a pair $\left( X,\left\{ \ell
_{k}\right\} _{k\in%
\mathbb{N}
}\right) $ where:

\begin{itemize}
\item $X$ is a graded vector space%
\begin{equation}
X=%
{\displaystyle\bigoplus_{n\in\mathbb{\mathbb{Z}}}}
X_{n}.
\end{equation}
Given an element $x\in X_{n}$ we say that $x$ has degree $n$, $\deg x=n$.

\item $\left\{ \ell_{k}\right\} _{k\in%
\mathbb{N}
}$ is a set of $k$-linear products of degree $k-2$ defined on $X$, i.e.,
given an arbitrary set of elements $x_{1},\ldots,x_{k}\in X$%
\begin{equation}
\deg\ell_{k}\left( x_{1},\ldots,x_{k}\right) =k-2+\deg x_{1}+\cdots+\deg
x_{k}.
\end{equation}

\item The products are graded symmetric%
\begin{equation}
\ell_{k}\left( x_{1},\ldots,x_{k}\right) =\left( -1\right) ^{\sigma
}\epsilon\left( \sigma,x\right) \ell_{k}\left( x_{\sigma\left( 1\right)
},\ldots,x_{\sigma\left( k\right) }\right) ,   \label{symml}
\end{equation}
where $\epsilon\left( \sigma,x\right) $ is the Kozul sign defined by a
graded symmetric product $x\wedge y=\left( -1\right) ^{\deg x\deg y}y\wedge x
$, which depends on the order of the elements $x_{1},\ldots,x_{k}$ and the
order of the permutation by means of the relation%
\begin{equation}
x_{1}\wedge\cdots\wedge x_{k}=\epsilon\left( \sigma,x\right) x_{\sigma
\left( 1\right) }\wedge\cdots\wedge x_{\sigma\left( k\right) }.
\end{equation}

\item The products $\ell_{k}$ satisfy the so-called $L_{\infty}$ identities%
\begin{equation}
\sum_{i+j=n+1}\left( -1\right) ^{i\left( j-1\right) }\sum_{\sigma\in
U_{n}}\left( -1\right) ^{\sigma}\epsilon\left( \sigma,x\right) \ell
_{j}\left( \ell_{i}\left( x_{\sigma\left( 1\right) },\ldots,x_{\sigma \left(
i\right) }\right) ,x_{\sigma\left( i+1\right) },\ldots ,x_{\sigma\left(
n\right) }\right) =0,   \label{Lid}
\end{equation}
with $n\geq1$, and $U_{n}$ being the set of unshuffle permutations of $n$
elements, i.e., permutation whose arguments satisfying the following
ordering relations%
\begin{align}
\sigma\left( 1\right) & <\cdots<\sigma\left( i\right) , \\
\sigma\left( i+1\right) & <\cdots<\sigma\left( n\right) .
\end{align}
\end{itemize}

\subsection{$L_{\infty}$ formulation of gauge theories}

Let us now focus on writing an arbitrary classical gauge theory in terms of
an $L_{\infty}$ algebra (see \cite{Zwie}). We begin by introducing a gauge theory with
fundamental field $A$ evaluated on a vector space $X_{-1}$. We also
introduce gauge transformations through a gauge parameter $\varepsilon$
taking values on another vector space $X_{0}$. The dynamics is determined by
the equation of motion $\mathcal{F}=0$, where the off-shell functions $%
\mathcal{F}$ take values on a vector space $X_{-2}$. It is possible to
define an $L_{\infty}$ algebra on $X=X_{0}\oplus X_{-1}\oplus X_{-2}$ such
that the whole information of the gauge theory is codified on some
non-vanishing $L_{\infty}$ products that satisfy the $L_{\infty}$ identities (\ref%
{Lid}). It is necessary to include the information related to three aspects
of the theory, namely, its definition of gauge transformations, the closed
gauge algebra and the equations of motion.

\subsubsection{Gauge transformations}

Given an arbitrary $L_{\infty}$ algebra, a set of gauge fields $A\in X_{-1}$
and a set of parameters $\varepsilon\in X_{0}$, the gauge variatons are
defined in terms of the $L_{\infty}$ products as follows%
\begin{equation}
\delta_{\varepsilon}A=\sum_{n=0}^{\infty}\frac{\left( -1\right) ^{\frac{%
n\left( n-1\right) }{2}}}{n!}\ell_{n+1}\left( \varepsilon ,A^{n}\right) . 
\label{gt}
\end{equation}
There are also trivial gauge transformations, i.e., equations of motion
symmetries. A particular case that will be important later and whose
presence is due to the non-vanishing products of three of more elements is
given by the following transformation%
\begin{equation}
\delta_{\varepsilon_{1},\varepsilon_{2}}^{T}A=\sum_{n=0}^{\infty}\frac{%
\left( -1\right) ^{\frac{n\left( n-1\right) }{2}}}{n!}\ell_{n+3}\left(
\varepsilon_{1},\varepsilon_{2},\mathcal{F},A^{n}\right) .   \label{t-g}
\end{equation}
This is a gauge transformation that vanishes on-shell and depends on two
parameters. Note that they do not appear in the case of Lie algebras because
their products are bilinear.

\subsubsection{Equations of motion}

To define a gauge invariant action, it is necessary to introduce an inner
product $\left\langle ~,~\right\rangle _{L_{\infty}}$ on $X$, satisfying the
following invariance conditions: Given $n+1$ elements $x_{0},\ldots,x_{n}\in
X$,%
\begin{align}
\left\langle x_{0},\ell_{n}\left( x_{1},x_{2},\ldots,x_{n}\right)
\right\rangle _{L_{\infty}} & =\left( -1\right) ^{1+\deg x_{0}\deg
x_{1}}\left\langle x_{1},\ell_{n}\left( x_{0},x_{2},\ldots,x_{n}\right)
\right\rangle _{L_{\infty}}, \\
\left\langle x_{0},x_{1}\right\rangle _{L_{\infty}} & =\left( -1\right)
^{\deg x_{0}\deg x_{1}}\left\langle x_{1},x_{0}\right\rangle _{L_{\infty}}.
\end{align}
The gauge transformations defined on Eq. (\ref{gt}) leave the following
action invariant%
\begin{equation}
S=\sum_{n=1}^{\infty}\frac{\left( -1\right) ^{\frac{n\left( n-1\right) }{2}}%
}{\left( n+1\right) !}\left\langle A,\ell_{n}\left( A^{n}\right)
\right\rangle _{L_{\infty}}.
\end{equation}
Taking the field variation of this action one finds%
\begin{equation}
\delta S=\left\langle \delta A,\mathcal{F}\right\rangle _{L_{\infty}}, 
\label{F}
\end{equation}
with%
\begin{equation}
\mathcal{F}=\sum_{n=1}^{\infty}\frac{\left( -1\right) ^{\frac{n\left(
n-1\right) }{2}}}{n!}\ell_{n}\left( A^{n}\right) .   \label{F1}
\end{equation}
Assuming non-degeneracy of the inner product \cite{Zwie,boots}, this leads
to the equation of motion $\mathcal{F}=0$.

\subsubsection{Gauge algebra}

Given two parameters $\varepsilon_{1},\varepsilon_{2}\in X_{0}$, the
commutator of its corresponding gauge transformations can be written in
terms of two gauge transformations 
\begin{equation}
\left[ \delta_{\varepsilon_{2}},\delta_{\varepsilon_{1}}\right]
A=\delta_{\varepsilon_{3}}A+\delta_{\varepsilon_{1},\varepsilon_{2}}^{T}A, 
\label{ga0}
\end{equation}
where the parameter $\varepsilon_{3}\in X_{0}$ is given by%
\begin{equation}
\varepsilon_{3}=\sum_{n=0}^{\infty}\frac{\left( -1\right) ^{\frac{n\left(
n-1\right) }{2}}}{n!}\ell_{n+2}\left(
\varepsilon_{1},\varepsilon_{2},A^{n}\right) .   \label{ga1}
\end{equation}
The presence of a trivial gauge transformation on Eq. (\ref{ga0}) imposes
some conditions on the dynamics. Starting from the definition of gauge
transformations in an arbitrary theory, it is possible to find the allowed equation of motion by
inspection of the algebraic element $\mathcal{F}$ on the trivial
transformation. However, since every product on Eq. (\ref{t-g}) has three or
more elements, this is not an issue when dealing with gauge symmetries
described by bilinear products, such as in the case of Lie algebras.

From Eq. (\ref{gt}) we can see that, given a gauge theory, the definition of
gauge variations defines the products of the corresponding $L_{\infty}$
algebra that involves elements on $X_{0}$ and $X_{-1}$. In the same way, from
Eqs. (\ref{t-g}) and (\ref{ga1}) we can see that the gauge-algebra-relations
of a theory, provide the information about the $L_{\infty}$ products acting
on at least two elements on $X_{0}$. Finally, by inspection of the equation of motion (%
\ref{F}) we can see that the dynamics of the theory is contained in the
products acting on elements of $X_{-1}$. This procedure was used detailed explained and used to write down the $L_{\infty}$ algebras that describe Yang--Mills and
three-dimensional CS theories in Ref. \cite%
{Zwie}.

\section{Free differential algebras}

The Chevalley--Eilenberg cohomology algebras of Lie algebras can be
formulated in terms of extended Maurer--Cartan equations that enlarge the
mathematical structure described by the standard Maurer--Cartan equations
for Lie algebras, including not only the usual left-invariant one-forms but
higher-degree differential forms \cite{dauria1}. This allows to extend de gauge principle
and study gauge invariant theories with higher-degree differential forms
whose symmetry is described by enlarged algebraic structures. Let us
consider an arbitrary manifold $M$ and basis of differential forms $\left\{
\Theta^{A\left( p\right) }\right\} _{p=1}^{N}$ defined on $\Lambda^{p}\left(
M\right) $. Each index $A\left( p\right) $ runs over a different domain
depending on the label $p$, which also denotes the degree of the
differential form $\Theta^{A\left( p\right) }$. This allows to write a set
of Maurer--Cartan equations that generalizes the standard Maurer--Cartan
equations of Lie algebras, defining a mathematical structure known as FDA%
\begin{equation}
\mathrm{d}\Theta^{A\left( p\right) }+\sum_{n=1}^{N}\frac{1}{n}C_{B_{1}\left(
p_{1}\right) \cdots B_{n}\left( p_{n}\right) }^{A\left( p\right)
}\Theta^{B_{1}\left( p_{1}\right) }\wedge\cdots\wedge\Theta ^{B_{n}\left(
p_{n}\right) }=0.   \label{fda6}
\end{equation}
The coefficients $C_{B_{1}\left( p_{1}\right) \cdots B_{n}\left(
p_{n}\right) }^{A\left( p\right) }$ are called generalized structure
constants (natural generalizations of the usual structure constants of Lie
algebras) and have graded symmetry in their lower indices which is ruled by
the (anti)symmetric wedge product between the differential forms in Eq. (\ref%
{fda6}). The nilpotent condition $\mathrm{d}^{2}\Theta^{A\left( p\right) }=0$
leads to the following Jacobi identity \cite{castellaniB}%
\begin{equation}
\sum_{m,n=1}^{N}\frac{1}{m}C_{B_{1}\left( p_{1}\right) \cdots B_{n}\left(
p_{n}\right) }^{A\left( p\right) }C_{C_{1}\left( q_{1}\right) \cdots
C_{m}\left( q_{m}\right) }^{B_{1}\left( p_{1}\right) }\Theta^{C_{1}\left(
q_{1}\right) }\wedge\cdots\wedge\Theta^{C_{m}\left( q_{m}\right)
}\wedge\Theta^{B_{2}\left( p_{2}\right) }\wedge\cdots\wedge\Theta
^{B_{n}\left( p_{n}\right) }=0.
\end{equation}

\subsection{Dual formulation}

Let us consider now a dual formulation for FDAs through the introduction of
a set of products acting on a dual basis. We introduce a graded vector space 
$\bar{X}$, a basis set $\left\{ T_{A\left( p\right) }\right\} _{p=1}^{N}$
with $\deg_{\bar{X}}T_{A\left( p\right) }=p$ and a set of $n$-linear
products acting on $\bar{X}$ ($n\geq1$)%
\begin{equation}
\left[ T_{A_{1}\left( p_{1}\right) },\ldots,T_{A_{n}\left( p_{n}\right) }%
\right] _{n}\in\bar{X}.   \label{Tprod}
\end{equation}
The components $\left[ T_{A_{1}\left( p_{1}\right) },\ldots,T_{A_{n}\left(
p_{n}\right) }\right] _{n}^{A\left( p\right) }$ are choosen to be
proportional to the structure constants of an FDA%
\begin{equation}
\left[ T_{A_{1}\left( p_{1}\right) },\ldots,T_{A_{n}\left( p_{n}\right) }%
\right] _{n}^{A\left( p\right) }=\left( n-1\right) !C_{A_{1}\left(
p_{1}\right) \cdots A_{n}\left( p_{n}\right) }^{A\left( p\right) }, 
\label{T}
\end{equation}
and therefore, they satisfy the same graded-symmetry relation%
\begin{equation}
\left[ T_{A_{\sigma\left( 1\right) }\left( p_{\sigma\left( 1\right) }\right)
},\ldots,T_{A_{\sigma\left( n\right) }\left( p_{\sigma\left( n\right)
}\right) }\right] _{n}=\epsilon\left( \sigma,T\right) \left[ T_{A_{1}\left(
p_{1}\right) },\ldots,T_{A_{n}\left( p_{n}\right) }\right] _{n}, 
\label{symmb}
\end{equation}
which, for $n=2$, leads to the following rule 
\begin{equation}
\left[ T_{A\left( r\right) },T_{B\left( s\right) }\right] _{2}=\left(
-1\right) ^{rs}\left[ T_{B\left( s\right) },T_{A\left( r\right) }\right] .
\end{equation}
From this definition it follows that the product $\left[ T_{A_{1}\left(
p_{1}\right) },\ldots,T_{A_{n}\left( p_{n}\right) }\right] _{n}$ lies in the
subspace $\bar{X}_{p}$ with $p=p_{1}+\cdots+p_{n}-1$, otherwise, the corresponding
structure constants vanish. Since the new products are proportional to the
generalized structure constants of an FDA, it is possible to write down the
generalized Jacobi in terms of them%
\begin{multline}
\sum_{m,n=1}^{N}\frac{1}{m!\left( n-1\right) !}\left[ \left[ T_{C_{1}\left(
q_{1}\right) },\ldots,T_{C_{m}\left( q_{m}\right) }\right]
_{m},T_{B_{2}\left( p_{2}\right) }\ldots,T_{B_{n}\left( p_{n}\right) }\right]
_{n}^{A\left( p\right) }\Theta^{C_{1}\left( q_{1}\right)
}\wedge\cdots\wedge\Theta^{C_{m}\left( q_{m}\right) } \\
\wedge\Theta^{B_{2}\left( p_{2}\right) }\wedge\cdots\wedge\Theta
^{B_{n}\left( p_{n}\right) }=0,   \label{Jac2}
\end{multline}
where the sum runs over the combinations of $p_{i}$ and $q_{i}$ such that $%
q_{1}+\cdots+q_{m}=p_{1}+1$ and $p_{1}+\cdots+p_{n}=p+1$. Since each element
on Eq. (\ref{Jac2}) is a power of order $m+n-1\,$\ in $\Theta$, we can
separate them in different equations, each one with the same power. By
renaming the indices and performing the sum over unsuffles we can sum on $%
m!\left( n-1\right) !$ equivalent elements. We then remove the factorial factors
and introduce a Kozul sign depending on the order of the permutation and the
degrees of the elements. The equation of power $l=m+n-1$ in $\Theta$ becomes 
\begin{equation}
\sum_{m+n=l-1}^{N}\sum_{\sigma\in U\left( l\right) }\epsilon\left(
\sigma,T\right) \left[ \left[ T_{B_{\sigma\left( 1\right) }\left(
q_{\sigma\left( 1\right) }\right) },\ldots,T_{B_{\sigma\left( m\right)
}\left( q_{\sigma\left( m\right) }\right) }\right] _{m},T_{B_{\sigma \left(
m+1\right) }\left( q_{\sigma\left( m+1\right) }\right)
}\ldots,T_{B_{\sigma\left( l\right) }\left( p_{\sigma\left( l\right)
}\right) }\right] _{n}=0,   \label{jacb}
\end{equation}
where we have also removed the dependence on the differential forms.

Now we define an $L_{\infty}$ algebra as follows. We introduce a new $%
\mathbb{Z}
$-graded vector space $X=\oplus_{n}X_{n}$, with a basis $\left\{ T_{A\left(
p\right) }\right\} _{p=1}^{N}$ such that $\deg_{X}T_{A\left( p\right) }=p-1$%
, endowed with the following set of products%
\begin{equation}
\ell_{n}\left( T_{A_{1}\left( p_{1}\right) },\ldots,T_{A_{n}\left(
p_{n}\right) }\right) =\left( -1\right) ^{\left( p_{1}-1\right) \left(
n-1\right) +\left( p_{2}-1\right) \left( n-2\right) +\cdots+\left(
p_{n-1}-1\right) }\left[ T_{A_{1}\left( p_{1}\right) },\ldots
,T_{A_{n}\left( p_{n}\right) }\right] _{n}.   \label{ln2}
\end{equation}
By replacing Eq. (\ref{ln2}) into Eqs. (\ref{jacb}) and (\ref{symmb}) it is
direct to prove that the $L_{\infty}$ products satisfy graded symmetry
relations and the $L_{\infty}$ identities from Eqs. (\ref{symml}) and (\ref%
{Lid}), showing that FDAs are dual to $L_{\infty}$ algebras. This could be
anticipated by noting that Eqs. (\ref{jacb}) are the $L_{\infty}$ identities
in the so called $b$-picture and Eq. (\ref{ln2}) is indeed the maping
between both equivalent formulations (see Ref. \cite{Zwie} for details on the $b$-picture). Here we make explicit the relation
between FDAs and $L_{\infty}$ algebras in the $\ell$-picture because it will
be necessary for the following sections. However, seeing such relation can be more easily found in the $b$-picture. For extensive reviews on the relation between $%
L_{\infty}$ algebras and graded differential algebras, see Refs. \cite%
{lada,lada2,kaj}.

As an example, let us consider an FDA carrying a one form $\Theta^{A\left(
1\right) }$ and a two-form $\Theta^{A\left( 2\right) }$. Such FDA is
described by the following Maurer--Cartan equations%
\begin{equation}
0=\text{d}\Theta^{A\left( 1\right) }+\frac{1}{2}C_{B\left( 1\right) C\left(
1\right) }^{A\left( 1\right) }\Theta^{B\left( 1\right)
}\wedge\Theta^{C\left( 1\right) },   \label{mc4}
\end{equation}%
\begin{align}
0 & =\text{d}\Theta^{A\left( 2\right) }+\frac{1}{2}\left( C_{B\left(
2\right) C\left( 1\right) }^{A\left( 2\right) }\Theta^{B\left( 2\right)
}\Theta^{C\left( 1\right) }+C_{B\left( 1\right) C\left( 2\right) }^{A\left(
2\right) }\Theta^{B\left( 1\right) }\Theta^{C\left( 2\right) }\right)  \notag
\\
& +\frac{1}{3}C_{B\left( 1\right) C\left( 1\right) D\left( 1\right)
}^{A\left( 2\right) }\Theta^{B\left( 1\right) }\wedge\Theta^{C\left(
1\right) }\wedge\Theta^{D\left( 1\right) }.   \label{mc5}
\end{align}
Eq. (\ref{mc4}) defines a Lie algebra with the antisymmetric structure
constants $C_{B\left( 1\right) C\left( 1\right) }^{A\left( 1\right) }$. On
the other hand, from Eqs. (\ref{T}) and (\ref{symmb}), the structure
constants $C_{B\left( 2\right) C\left( 1\right) }^{A\left( 2\right) }$ are
symmetric in the lower indices, allowing us to sum both terms in the
right-hand side of (\ref{mc5}) while the $3$-cocycle $C_{B\left( 1\right)
C\left( 1\right) D\left( 1\right) }^{A\left( 2\right) }$ is completely
antisymmetric. We now consider the two-form in the adjoint representation of
the Lie algebra. This means that both indices $A\left( 1\right) $ and $%
A\left( 2\right) $ take the same values. We denote $A\left( 1\right)
=A\left( 2\right) =A$, making necessary to rename the FDA-potentials as $%
\Theta^{A\left( 1\right) }=\Theta_{1}^{A}$ and $\Theta^{A\left( 2\right)
}=\Theta_{2}^{A}$. The structure constants of the FDA can be simply writen as%
\begin{align}
C_{B\left( 1\right) C\left( 1\right) }^{A\left( 1\right) } & =\left[
T_{B\left( 1\right) },T_{C\left( 1\right) }\right] _{2}^{A\left( 1\right)
}=C_{BC}^{A},  \label{c1} \\
C_{B\left( 2\right) C\left( 1\right) }^{A\left( 2\right) } & =\left[
T_{B\left( 2\right) },T_{C\left( 1\right) }\right] _{2}^{A\left( 2\right)
}=C_{BC}^{A},  \label{c2} \\
C_{C\left( 1\right) B\left( 2\right) }^{A\left( 2\right) } & =\left[
T_{C\left( 1\right) },T_{B\left( 2\right) }\right] _{2}^{A\left( 2\right)
}=-C_{CB}^{A},  \label{c3} \\
C_{B\left( 1\right) C\left( 1\right) D\left( 1\right) }^{A\left( 2\right) }
& =\frac{1}{2}\left[ T_{B\left( 1\right) },T_{C\left( 1\right) },T_{D\left(
1\right) }\right] _{3}^{A\left( 2\right) }=C_{BCD}^{A},   \label{c4}
\end{align}
where $C_{BC}^{A}$ are the antisymmetric structure constants of the original
Lie algebra. Note that we are using an antisymmetric tensor $C_{BC}^{A}$ to
write down the components of a symmetric tensor $C_{B\left( 2\right) C\left(
1\right) }^{A\left( 2\right) }$. Thus, the Maurer--Cartan equations become%
\begin{align}
0 & =\mathrm{d}\Theta_{1}^{A}+\frac{1}{2}C_{BC}^{A}\Theta_{1}^{B}\wedge%
\Theta_{1}^{C},  \label{mc3b} \\
0 & =\mathrm{d}\Theta_{2}^{A}+C_{BC}^{A}\Theta_{2}^{B}\wedge\Theta_{1}^{C}+%
\frac{1}{3}C_{BCD}^{A}\Theta_{1}^{B}\wedge\Theta_{1}^{C}\wedge\Theta
_{1}^{D}.   \label{mc3}
\end{align}
We can write this FDA as an $L_{\infty}$ algebra in the $\ell$-picture. We
define the graded vector space $X=X_{0}\oplus X_{1}$ with basis $\left\{
T_{A},\tilde{T}_{A}\right\} $ (with $T_{A}\in X_{0}$ and $\tilde{T}_{A}\in
X_{1}$) equipped with the following products%
\begin{align}
\ell_{2}\left( T_{B},T_{C}\right) & =C_{BC}^{A}T_{A},  \label{l1a} \\
\ell_{2}\left( T_{B},\tilde{T}_{C}\right) & =-C_{BC}^{A}\tilde{T}_{A},
\label{l1b} \\
\ell_{3}\left( T_{B},T_{C},T_{D}\right) & =2C_{BCD}^{A}\tilde{T}_{A},
\label{l1c} \\
\text{Others} & =0.   \label{l1d}
\end{align}
Eqs. (\ref{l1a}) - (\ref{l1c}) carry the information of the structure
constants from Eqs. (\ref{c1}) - (\ref{c4}) in $L_{\infty}$ formalism with $%
T_{A\left( 1\right) }=T_{A}$ and $T_{A\left( 2\right) }=\tilde{T}_{A}$. As
we have seen in Eq. (\ref{jacb}), those products satisfy the $L_{\infty}$
properties. This example explicitly shows how the original FDA-product $%
\left[ \Theta^{\left( 2\right) },\Theta^{\left( 1\right) }\right] ^{A\left(
2\right) }=C_{B\left( 2\right) C\left( 1\right) }^{A\left( 2\right)
}\Theta^{B\left( 2\right) }\wedge\Theta^{C\left( 1\right) }$ is symmetric
because of the symmetry of the structure constants in the lower indices $%
C_{B\left( 2\right) C\left( 1\right) }^{A\left( 2\right) }=C_{C\left(
1\right) B\left( 2\right) }^{A\left( 2\right) }$. However, these constants
are builded with the antisymmetric structure constants of a Lie algebra.
Moreover, we can also note that by imposing $C_{BCD}^{A}=0$, the resulting $%
L_{\infty}$ algebra becomes into a trivial enlargement of the Lie algebra on
Eqs. (\ref{l1a}) and (\ref{l1b}) due to the absence of a non-trivial cocycle
providing additional structure.

\subsection{Gauged FDAs}

From now on, we will focus in the particular FDA having a one-form $A^{A}$
and a $p$-form $A^{i}$. In such algebra, known as FDA1 and extensively
studied in Refs. \cite{cast1,cast2,cast3,cast4}, Eq. (\ref{fda6}) is reduced
to two Maurer--Cartan equations\footnote{%
From now on, we will omit the wedge product between differential forms.}:%
\begin{align}
\mathrm{d}A^{A}+\frac{1}{2}C_{BC}^{A}A^{B}A^{C} & =R^{A}=0,  \label{mu} \\
\mathrm{d}A^{i}+C_{Aj}^{i}A^{A}A^{j}+\frac{1}{\left( p+1\right) !}%
C_{A_{1}\cdots A_{p+1}}^{i}A^{A_{1}}\cdots A^{A_{p+1}} & =R^{i}=0. 
\label{B}
\end{align}

We introduce gauge variations in terms of the standard and generalized
covariant derivatives of a set of parameters. Let $\varepsilon^{A}$ and $%
\varepsilon^{i}$ be $0$-form and a $\left( p-1\right) $-form gauge
parameters respectively. The corresponding gauge variations are defined as
follows 
\begin{align}
\delta A^{A} & =\mathrm{d}\varepsilon^{A}+C_{BC}^{A}A^{B}\varepsilon ^{C},
\label{d1} \\
\delta A^{i} & =\mathrm{d}\varepsilon^{i}+C_{Aj}^{i}A^{A}\varepsilon
^{j}-C_{Aj}^{i}\varepsilon^{A}A^{j}-\frac{1}{p!}C_{A_{1}\cdots
A_{p+1}}^{i}\varepsilon^{A_{1}}A^{A_{2}}\cdots A^{A_{p+1}}.   \label{d2}
\end{align}
Eq. (\ref{d1}) corresponds to the usual gauge transformation of a one-form
and Eq. (\ref{d2}) extends the definition of covariant derivative to a $%
\left( p-1\right) $-form involving the $0$-form parameter and the new
structure constants.

With those transformations defined, it is possible to write down a gauge
invariant theory. The definition of transgression and CS forms for FDA1 can
be found by studying the corresponding Chern--Weil theorem \cite{CSFDA,cw,tf}.
Let $A=\left( A^{A},A^{i}\right) $ be a set of gauge fields composed by a
one-form and a $p$-form. Let $R=\left( R^{A},R^{i}\right) $ be its
corresponding field strength whose components are the standard and
generalized non-vanishing curvatures defined by gauging of the
Maurer--Cartan equations (\ref{mu}) and (\ref{B}). Using combinations of the
components of $R$, it is possible to define an invariant $q$-form as follows%
\begin{equation}
\chi_{q}\left( A\right) =\sum_{m,n}g_{A_{1}\cdots A_{m}i_{1}\cdots
i_{n}}R^{A_{1}}\cdots R^{A_{m}}R^{i_{1}}\cdots R^{i_{n}},
\end{equation}
where the sum runs over all the possible combinations such that $2m+\left(
p+1\right) n=q$. In order to $\chi_{q}$ be gauge invariant, the coefficients 
$g_{A_{1}\cdots A_{m}i_{1}\cdots i_{n}}$ must satisfy the following
conditions%
\begin{align}
\sum_{r=1}^{m}C_{A_{0}A_{r}}^{C}g_{A_{1}\cdots\hat{A}_{r}C\cdots
A_{m}i_{1}\cdots i_{n}}+\sum_{s=1}^{n}C_{A_{0}i_{s}}^{k}g_{A_{1}\cdots
A_{m}i_{1}\cdots\hat{\imath}_{s}k\cdots i_{n}} & =0,  \label{it1} \\
\sum_{r=1}^{m+1}C_{A_{r}B_{1}\cdots B_{p}}^{i_{1}}g_{A_{1}\cdots\hat{A}%
_{r}\cdots A_{m+1}i_{1}\cdots i_{n}} & =0,  \label{it2} \\
\sum_{r=1}^{m+1}C_{A_{r}j}^{i_{1}}g_{A_{1}\cdots\hat{A}_{r}\cdots
A_{m+1}i_{1}\cdots i_{n}} & =0,   \label{it3}
\end{align}
where indices with hat denote the absence of such indices. Eqs. (\ref%
{it1}) - (\ref{it3}) are the corresponding generalization of the invariant
tensor conditions for Lie algebras. The generalized invariant tensor
conditions also make $\chi_{q}$ closed.

Let us introduce a second set of gauge fields and field strengths $\bar {A}%
=\left( \bar{A}^{A},\bar{A}^{i}\right) $ and $\bar{R}=\left( \bar{R}^{A},%
\bar{R}^{i}\right) $. It also is possible to write down an invariant $q$%
-form $\chi_{q}\left( \bar{A}\right) $ using only the components of $\bar{R}$
as building blocks. The difference between both invariant $q$-forms is given
by the exterior derivative of a $\left( q-1\right) $-form, as follows%
\begin{equation}
\chi_{q}\left( A\right) -\chi_{q}\left( \bar{A}\right) =\mathrm{d}%
Q_{q-1}\left( A,\bar{A}\right) .   \label{cw1}
\end{equation}
The $\left( q-1\right) $-form $Q_{q-1}\left( A,\bar{A}\right) $ is called
transgression form and is explicitly given by%
\begin{equation}
Q_{q-1}\left( A,\bar{A}\right) =\sum_{m,n}g_{A_{1}\cdots A_{m}i_{1}\cdots
i_{n}}\int_{0}^{1}\mathrm{d}t\left( mu^{A_{1}}R_{t}^{A_{2}}\cdots
R_{t}^{A_{m}}R_{t}^{i_{1}}\cdots R_{t}^{i_{n}}+nR_{t}^{A_{1}}\cdots
R_{t}^{A_{m}}u^{i_{1}}R_{t}^{i_{2}}\cdots R_{t}^{i_{n}}\right) , 
\label{cw2}
\end{equation}
where we introduce an homotopic gauge field $A_{t}=\bar{A}+t\left( A-\bar {A}%
\right) =\left( A_{t}^{A},A_{t}^{i}\right) $ and its corresponding field
strength $R_{t}=\left( R_{t}^{A},R_{t}^{i}\right) $. By locally setting $%
\bar{A}=0$, Eq. (\ref{cw2}) becomes a definion of $\left( q-1\right) $%
-dimensional CS form invariant under the transformations of FDA1. Moreover,
by imposing $n=0$, the generalized invariant tensor conditions (\ref{mu})
reproduce the standard one for Lie algebras. In the same way, Eqs. (\ref{cw1}%
) and (\ref{cw2}) reproduce the standard Chern--Weil theorem and
transgression forms for Lie algebras.

\subsection{Dynamics}

From now on, we will use a compact notation for FDA-valued algebraic
elements. The details of this notation can be found in Appendix A. Let us
then consider a transgression action carrying one $p$-form extension defined
on a $q-1$ dimensional manifold $M_{q-1}$%
\begin{equation}
S_{\mathrm{T}}=\int_{M_{q-1}}\sum_{m,n}\int_{0}^{1}\mathrm{d}t\left(
m\left\langle u,R_{t}^{m-1};R_{t}^{n}\right\rangle +n\left\langle
R_{t}^{m};u,R_{t}^{n-1}\right\rangle \right) .   \label{st}
\end{equation}
By taking the total variation of $S_{\text{T}}$ and integrating by parts we
find%
\begin{align}
\delta S_{\mathrm{T}} & =\int_{M_{q-1}}\sum_{m,n}\int_{0}^{1}\mathrm{d}%
t\left( m\left\langle \delta u,R_{t}^{m-1};R_{t}^{n}\right\rangle +m\left(
m-1\right) \left\langle \nabla_{t}u,\delta
A_{t},R_{t}^{m-2};R_{t}^{n}\right\rangle \right.  \notag \\
& +mn\left\langle \nabla_{t}u,R_{t}^{m-1};\delta
A_{t},R_{t}^{n-1}\right\rangle +mn\left\langle \delta
A_{t},R_{t}^{m-1};\nabla_{t}u,R_{t}^{n-1}\right\rangle +n\left\langle
R_{t}^{m};\delta u,R_{t}^{n-1}\right\rangle  \notag \\
& \left. -\left( -1\right) ^{p}n\left( n-1\right) \left\langle
R_{t}^{m};\nabla_{t}u,\delta A_{t},R_{t}^{n-2}\right\rangle \right) +\text{%
Boundary terms.}   \label{deltaS}
\end{align}
Using the definition of homotopic gauge fields and curvatures, we find the
following relations

\begin{equation}
\frac{\mathrm{d}R_{t}}{\mathrm{d}t}=\nabla_{t}u,\text{ \ \ \ }\frac {\mathrm{%
d}\delta A_{t}}{\mathrm{d}t}=\delta u.   \label{Rels}
\end{equation}
By plugging in Eqs. (\ref{Rels}) into Eq. (\ref{deltaS}), integrating by
parts with respect to the parameter $t$ and neglecting the boundary terms,
the variation of the action takes the form%
\begin{equation}
\delta S_{\mathrm{T}}=\int_{M_{q-1}}\sum_{m,n}\left( m\left\langle \delta
A,R^{m-1};R^{n}\right\rangle +n\left\langle R^{m};\delta
A,R^{n-1}\right\rangle -m\left\langle \delta\bar{A},\bar{R}^{m-1};\bar{R}%
^{n}\right\rangle -n\left\langle \bar{R}_{t}^{m};\delta\bar{A},\bar{R}%
^{n-1}\right\rangle \right) .
\end{equation}
Since the variations of the gauge fields $A$ and $\bar{A}$ are independent,
we obtain two equations of motion:%
\begin{align}
\sum_{m,n}\left( m\left\langle \delta A,R^{m-1};R^{n}\right\rangle
+n\left\langle R^{m};\delta A,R^{n-1}\right\rangle \right) & =0,
\label{teom1} \\
\sum_{m,n}\left( m\left\langle \delta\bar{A},\bar{R}^{m-1};\bar{R}%
^{n}\right\rangle +n\left\langle \bar{R}_{t}^{m};\delta\bar{A},\bar{R}%
^{n-1}\right\rangle \right) & =0.   \label{teoom2}
\end{align}
Those equations are reduced equations of motion for a standard transgression
action if we impose $n=0$. Moreover, by locally setting $\bar{A}%
^{A}=0$ and $\bar{A}^{i}=0$ we obtain the equations of motion for extended
CS theory (or FDA1-CS theory) that can be separated again on its independent
variations with respect to the~one-form and $p$-form as follows%
\begin{align}
\delta A^{A} & :\text{ \ \ }\sum_{m,n}mg_{A_{1}A_{2}\cdots A_{m}i_{1}\cdots
i_{n}}R^{A_{2}}\cdots R^{A_{m}}R^{i_{1}}\cdots R^{i_{n}}=0,  \label{cseom1}
\\
\delta A^{i} & :\text{ \ \ }\sum_{m,n}ng_{A_{1}\cdots A_{m}i_{1}i_{2}\cdots
i_{n}}R^{A_{1}}\cdots R^{A_{m}}R^{i_{2}}\cdots R^{i_{n}}=0.   \label{cseom2}
\end{align}

\section{$L_{\infty}$ formulation of CS theory}

In this section we will follow the procedure introduced on Ref. \cite{Zwie}
to find the $L_{\infty}$ structure of standard $2m-1$ dimensional CS theory,
whose action can be found from Eq. (\ref{st}) by setting $\bar{A}=0$ in
absence of $p$-form gauge fields, i.e.,\footnote{%
We include a label to denote the invariant tensor of the Lie algebra and
distinguish it from the inner product of $L_{\infty}$ algebras. We will
later introduce an FDA1 invariant tensor that can be identified by a
semicolon separating the algebraic sectors.} 
\begin{equation}
S_{\mathrm{CS}}=m\int_{M_{2m-1}}\int_{0}^{1}\mathrm{d}t\left\langle
A,R_{t}^{m-1}\right\rangle _{\text{\textrm{Lie}}},   \label{cs}
\end{equation}
with $A_{t}=tA$ and $R_{t}=\left( t^{2}-t\right) R$. By comparing with the
general formulation of gauge theories in terms of $L_{\infty}$ algebras, we
will extract the relevant information of the theory contained in the gauge
transformations, their closed gauge algebra and the equations of motion, and will
write it in terms of algebraic products.

\subsection{Gauge transformations}

In standard CS theory, the fundamental field is the one-form gauge
field $A_{\mu}^{A}$ whose gauge variation is given by the Lie-covariant
derivative of a $0$-form gauge parameter $\varepsilon^{A}$%
\begin{equation}
\delta
A_{\mu}^{A}=\partial_{\mu}\varepsilon^{A}+C_{BC}^{A}A_{\mu}^{B}%
\varepsilon^{C}.   \label{Lie}
\end{equation}
We identify the parameters $\varepsilon^{A}\in X_{0}$. In the $L_{\infty}$
formulation of gauge theories, the gauge variation of $A_{\mu}^{A}$ can be
written in terms of the $L_{\infty}$ products according to Eq. (\ref{gt}).
By direct inspection of Eq. (\ref{Lie}) we can remove every term in Eq. (\ref%
{gt}) except by the ones that are powers of degree zero and one in the gauge
field. This can be written in components as%
\begin{equation}
\delta A_{\mu}^{A}=\left[ \ell_{1}\left( \varepsilon\right) \right] _{\mu
}^{A}+\left[ \ell_{2}\left( \varepsilon,A\right) \right] _{\mu}^{A},
\end{equation}
which leads to the following information about the $L_{\infty}$ products:%
\begin{align}
\left[ \ell_{1}\left( \varepsilon\right) \right] _{\mu}^{A} &
=\partial_{\mu}\varepsilon^{A},  \label{tr1} \\
\left[ \ell_{2}\left( \varepsilon,A\right) \right] _{\mu}^{A} & =\left[
A_{\mu},\varepsilon\right] ^{A}.   \label{tr2}
\end{align}
Any other product involving one element on $X_{0}$ and elements from $X_{-1}$ vanishes.

\subsection{Gauge algebra}

As second step, we need to ensure the closure of the gauge subalgebra $%
L_{\infty}^{\mathrm{gauge}}$. The commutator between two gauge
transformations can be written in terms of a third gauge transformation
without introducing on-shell symmetries%
\begin{equation}
\left( \delta_{2}\delta_{1}-\delta_{1}\delta_{2}\right)
A_{\mu}^{A}=\partial_{\mu}\varepsilon_{3}^{A}+C_{DB}^{A}A_{\mu}^{D}%
\varepsilon_{3}^{B}.   \label{dd1}
\end{equation}
The components of the new gauge parameter $\varepsilon_{3}$ are given by the
Lie product between the original parameters, i.e., $%
\varepsilon_{3}^{A}=C_{BC}^{A}\varepsilon_{2}^{B}\varepsilon_{1}^{C}$. On
the other hand, from Eqs. (\ref{ga0}) and (\ref{ga1}) we get%
\begin{equation}
\left( \delta_{2}\delta_{1}-\delta_{1}\delta_{2}\right) A_{\mu}^{A}=\left[
\ell_{1}\left( \ell_{2}\left( \varepsilon_{1},\varepsilon_{2}\right) \right) %
\right] _{\mu}^{A}+\left[ \ell_{1}\left( \ell_{3}\left(
\varepsilon_{1},\varepsilon_{2},A\right) \right) \right] _{\mu}^{A}+\left[
\ell_{2}\left( \ell_{2}\left( \varepsilon_{1},\varepsilon_{2}\right)
,A\right) \right] _{\mu}^{A}.   \label{dd2}
\end{equation}
We can find other products that, for consistency, must be non-vanishing.
Since $\ell_{2}\left( \varepsilon_{1},\varepsilon_{2}\right) ,\ell_{3}\left(
\varepsilon_{1},\varepsilon_{2},A\right) \in X_{0}$ we get%
\begin{align}
\left[ \ell_{1}\left( \ell_{2}\left( \varepsilon_{1},\varepsilon _{2}\right)
\right) \right] _{\mu}^{A} & =\partial_{\mu}\left[ \ell _{2}\left(
\varepsilon_{1},\varepsilon_{2}\right) \right] ^{A}, \\
\left[ \ell_{1}\left( \ell_{3}\left( \varepsilon_{1},\varepsilon
_{2},A\right) \right) \right] _{\mu}^{A} & =\partial_{\mu}\left[
\ell_{3}\left( \varepsilon_{1},\varepsilon_{2},A\right) \right] ^{A},
\end{align}
and therefore, by comparing Eqs. (\ref{dd1}) and (\ref{dd2}), we obtain the
following products between elements in $X_{0}$%
\begin{equation}
\left[ \ell_{2}\left( \varepsilon_{1},\varepsilon_{2}\right) \right]
^{A}=C_{BC}^{A}\varepsilon_{2}^{B}\varepsilon_{1}^{C},\text{ \ \ \ }\left[
\ell_{3}\left( \varepsilon_{1},\varepsilon_{2},A\right) \right] _{\mu}^{A}=0.
\end{equation}

In summary, at this point the information is codified into the following $L_{\infty}$ products%
\begin{equation}
\begin{tabular}{ccc}
$%
\begin{array}{l}
\left[ \ell _{1}\left( \varepsilon \right) \right] _{\mu }^{A}=\partial
_{\mu }\varepsilon ^{A}, \\ 
\left[ \ell _{2}\left( \varepsilon ,A\right) \right] _{\mu }^{A}=\left[
A_{\mu },\varepsilon \right] ^{A},%
\end{array}%
$ &  & $\left[ \ell _{2}\left( \varepsilon _{1},\varepsilon _{2}\right) %
\right] ^{A}=\left[ \varepsilon _{2},\varepsilon _{1}\right] ^{A}.$ \\ 
Gauge transformations &  & G$\text{auge algebra}$%
\end{tabular}%
\ \ \ \ \ \ \ \text{ }  \label{gauge}
\end{equation}%
Any other product involving elements on $X_{0}$ and $X_{-1}$ vanishes. Those
products define an $L_{\infty }$ algebra and describe a consistent gauge
theory if we include the information concerning the dynamics.

\subsection{Equations of motion}

Starting from the $2m-1$ dimensional CS action (\ref{cs}), one finds the
following variation%
\begin{equation}
\delta S_{\mathrm{CS}}=\int_{M_{2m-1}}\left\langle \delta
A,R^{m-1}\right\rangle _{\mathrm{Lie}}.
\end{equation}%
Expanding the power of the curvature, it is possible to write the variation
of $S_{\text{CS}}$ as 
\begin{equation}
\delta S_{\mathrm{CS}}=\int_{M_{2m-1}}\sum_{k=0}^{m-1}\binom{m-1}{k}%
\left\langle \delta A,\left( \mathrm{d}A\right) ^{m-k-1},\frac{1}{2^{k}}%
\left[ A,A\right] ^{k}\right\rangle _{\mathrm{Lie}},
\end{equation}%
or, explicitly writing algebraic indices and component of the differential
forms:%
\begin{align}
\delta S_{\mathrm{CS}}& =\int \mathrm{d}x^{2m-1}\sum_{k=0}^{m-1}\frac{1}{%
2^{k}}\binom{m-1}{k}\varepsilon ^{\mu _{1}\cdots \mu _{2m-1}}g_{AB_{1}\cdots
B_{m-1}}\delta A_{\mu _{1}}^{A}\partial _{\mu _{2}}A_{\mu
_{3}}^{B_{1}},\cdots \partial _{\mu _{2m-2k-3}}A_{\mu _{m-k+1}}^{B_{m-k-1}} 
\notag \\
& \times \left[ A_{\mu _{2m-2k-1}},A_{\mu _{2m-2k}}\right] ^{B_{m-k}}\cdots %
\left[ A_{\mu _{2m-2}},A_{\mu _{2m-1}}\right] ^{B_{m-1}},  \label{delta1}
\end{align}%
where $\varepsilon ^{\mu _{1}\cdots \mu _{2m-1}}$ is the Levi--Civita
pseudotensor and $g_{B_{1}\cdots B_{m}}$ are the components of the invariant
tensor of the Lie algebra. Note that the invariant tensor is given by the
trace over the Lie algebra's basis and it can be understood as a multilinear
product. On the other hand, the inner product of the $L_{\infty }$ algebra
is bilinear. By comparing Eqs. (\ref{F}) and (\ref{delta1}) we can idetify
the inner product of the $L_{\infty }$ algebra in terms of the invariant
tensor of the Lie algebra. Given two algebraic elements $x\in X_{-1}$ and $%
y\in X_{-2}$ evaluated on the Lie algebra, we identify%
\begin{equation}
\left\langle x,y\right\rangle _{L_{\infty }}=\int \mathrm{d}x^{2m-1}\eta
^{\mu \nu }\left\langle x_{\mu },y_{\nu }\right\rangle _{\mathrm{Lie}}.
\end{equation}%
Then, the variation of the action can be written in terms of $\mathcal{F}$
as follows: 
\begin{align}
\delta S_{\mathrm{CS}}& =\left\langle \delta A,\mathcal{F}\right\rangle
_{L_{\infty }}  \notag \\
& =\int \mathrm{d}x^{2m-1}\eta ^{\mu \nu }\left\langle \delta A_{\mu },%
\mathcal{F}_{\nu }\right\rangle _{\mathrm{Lie}}.
\end{align}%
In this case, the algebraic index is the Lie algebra index. Therefore we can
write%
\begin{equation}
\left\langle \delta A,\mathcal{F}\right\rangle _{L_{\infty }}=\int \mathrm{d}%
x^{2m-1}\eta ^{\mu \nu }g_{AB}\delta A_{\mu }^{A}\mathcal{F}_{\nu }^{B},
\label{delta2}
\end{equation}%
where $g_{AB}$ is the Cartan--Killing metric of the Lie algebra. By
comparing Eqs. (\ref{delta1}) and (\ref{delta2}) we obtain an explicit
expression for $\mathcal{F}$, namely 
\begin{align}
\mathcal{F}_{\nu }^{A}& =\sum_{k=0}^{m-1}\frac{1}{2^{k}}\binom{m-1}{k}%
\varepsilon _{\nu }^{\text{ \ }\mu _{1}\cdots \mu _{2m-2}}g_{\text{ \ }%
B_{1}\cdots B_{m-1}}^{A}\partial _{\mu _{1}}A_{\mu _{2}}^{B_{1}}\cdots
\partial _{\mu _{2m-2k-3}}A_{\mu _{2m-2k-2}}^{B_{m-k-1}}  \notag \\
& \times \left[ A_{\mu _{2m-2k-1}},A_{\mu _{2m-2k}}\right] ^{B_{m-k}}\cdots %
\left[ A_{\mu _{2m-3}},A_{\mu _{2m-2}}\right] ^{B_{m-1}}.  \label{FCS1}
\end{align}

On the other hand, the algebraic element $\mathcal{F}$, in general given by
Eq. (\ref{F1}), can be written for the case as 
\begin{equation}
\mathcal{F}^{A}=\sum_{l=1}^{\infty }\frac{\left( -1\right) ^{\frac{l\left(
l-1\right) }{2}}}{l!}\left[ \ell _{l}\left( A^{l}\right) \right] ^{A}.
\label{FCS2}
\end{equation}%
We can now extract the corresponding information about the $L_{\infty }$
products. From Eqs. (\ref{FCS1}) and (\ref{FCS2}) we have 
\begin{align}
\sum_{l=1}^{\infty }\frac{\left( -1\right) ^{\frac{l\left( l-1\right) }{2}}}{%
l!}\left[ \ell _{l}\left( A^{l}\right) \right] _{\nu }^{A}& =\sum_{k=0}^{m-1}%
\frac{1}{2^{k}}\binom{m-1}{k}\varepsilon _{\nu }^{\text{ \ }\mu _{1}\cdots
\mu _{2m-2}}g_{\text{ \ }B_{1}\cdots B_{m-1}}^{A}\partial _{\mu _{1}}A_{\mu
_{2}}^{B_{1}},\cdots\partial _{\mu _{2m-2k-3}}A_{\mu
_{2m-2k-2}}^{B_{m-k-1}}  \notag \\
& \times \left[ A_{\mu _{2m-2k-1}},A_{\mu _{2m-2k}}\right] ^{B_{m-k}}\cdots
\left[ A_{\mu _{2m-3}},A_{\mu _{2m-2}}\right] ^{B_{m-1}}.
\end{align}%
We compare then the terms of equal powers of $A$. The $k$-th element in the
sum of the right-hand-side has degree $m+k-1$ in $A$. There is only one
element of such degree. We can match one element on the left-hand side with
one element on the right side. Therefore, given a fixed value of $m$, we
have many values for $k$ ($k=0,\ldots ,m-1$), and given a value for $k$, $l$
is completely determined by%
\begin{equation}
l=m+k-1.
\end{equation}%
In other words, for a fixed value of $m$, we have a $2m-1$ dimensional gauge
theory whose dynamics is described by a set of non-vanishing products $\ell
_{l}$ with ($l=m-1,\ldots ,2m-2$) acting on $X_{-1}$. In general the $l$-th
product is given by%
\begin{align}
\left[ \ell _{l}\left( A^{l}\right) \right] _{\nu }^{A}& =\left( -1\right) ^{%
\frac{l\left( l-1\right) }{2}}\frac{1}{2^{l-m+1}}\frac{l!\left( m-1\right) !%
}{\left( 2m-l-2\right) !\left( l-m+1\right) !}\varepsilon _{\nu }^{\text{ \ }%
\mu _{1}\cdots \mu _{2m-2}}  \notag \\
& \times g_{\text{ \ }B_{1}\cdots B_{m-1}}^{A}\partial _{\mu _{1}}A_{\mu
_{2}}^{B_{1}}\cdots \partial _{\mu _{4m-2l-5}}A_{\mu _{4m-2l-4}}^{B_{2m-l-2}}
\notag \\
& \times \left[ A_{\mu _{4m-2l-3}},A_{\mu _{4m-2l-2}}\right]
^{B_{2m-l-1}}\cdots \left[ A_{\mu _{2m-3}},A_{\mu _{2m-2}}\right] ^{B_{m-1}}.
\label{ln-cs}
\end{align}

In summary, we can write the $2m+1$ dimensional CS theory (for convenience
we change $m\longrightarrow m+1$) as an $L_{\infty }$ algebra defined by a
vector space $X_{0}\oplus X_{-1}\oplus X_{-2}$ endowed with the following
products%
\begin{align}
\left[ \ell _{1}\left( \varepsilon \right) \right] _{\mu }^{A}& =\partial
_{\mu }\varepsilon ^{A},  \label{L1} \\
\left[ \ell _{2}\left( \varepsilon ,A\right) \right] _{\mu }^{A}& =\left[
A_{\mu },\varepsilon \right] ^{A},  \label{L2} \\
\left[ \ell _{2}\left( \varepsilon _{1},\varepsilon _{2}\right) \right]
^{A}& =\left[ \varepsilon _{2},\varepsilon _{1}\right] ^{A},  \label{L3} \\
\left[ \ell _{2}\left( \varepsilon ,E\right) \right] _{\nu }^{A}& =\left[
E_{\nu },\varepsilon \right] ^{A},  \label{LE}
\end{align}%
\begin{align}
\left[ \ell _{l}\left( A_{1},...,A_{l}\right) \right] _{\nu }^{A}& =\frac{1}{%
2^{l-m}}\frac{\left( -1\right) ^{\frac{N\left( N-1\right) }{2}}l!m!}{\left(
2m-l\right) !\left( l-m\right) !}\varepsilon _{\nu }^{\text{ \ }\mu
_{1}\cdots \mu _{2m}}g_{\text{ \ }B_{1}\cdots B_{m}}^{A}\partial _{\mu
_{1}}(A_{\{1})_{\mu _{2}}^{B_{1}}\cdots \partial _{\mu
_{4m-2l-1}}(A_{2m-l})_{\mu _{4m-2l}}^{B_{2m-l}}  \notag \\
& \times \left[ (A_{2m-l+1})_{\mu _{4m-2l+1}},(A_{2m-l+2})_{\mu _{4m-2l+2}}%
\right] ^{B_{2m-l+1}}\cdots \left[ (A_{l-1})_{\mu _{2m-1}},(A_{l\}})_{\mu
_{2m}}\right] ^{B_{m}},  \label{L4}
\end{align}%
with $\varepsilon ,\varepsilon _{1},\varepsilon _{2}\in X_{0}$, $%
A,A_{1},\ldots ,A_{l}\in X_{-1}$, $E\in X_{-2}$ and $l=m,\ldots ,2m$. Note
that Eq. (\ref{L4}) is obtained directly from Eq. (\ref{ln-cs}) by
considering arbitrary elements on $X_{-1}$ instead of the same gauge field $l
$ times. The corresponding symmetrization is included in order to ensure
that the symmetry rule of the $L_{\infty }$ products holds. Moreover, the
product involving elements on space $X_{-2}$ in Eq. (\ref{L4}) is obtained
by consistency with the $L_{\infty }$ identities (\ref{Lid}). The
calculation can be found in Appendix B.1.

As an example, let us consider the tree-dimensional CS theory. This case is
obtained by setting $m=1$. In such case there are only two products $\ell
_{l}$ ($l=1,2$) in the dynamical sector%
\begin{align}
\left[ \ell _{1}\left( A_{1}\right) \right] _{\nu }^{A}& =\varepsilon _{\nu
}^{\text{ \ }\mu _{1}\mu _{2}}\partial _{\mu _{1}}\left( A_{1}^{A}\right)
_{\mu _{2}},  \label{three1} \\
\left[ \ell _{2}\left( A_{1},A_{2}\right) \right] _{\nu }^{A}& =-\varepsilon
_{\nu }^{\text{ \ }\mu _{1}\mu _{2}}\left[ \left( A_{1}\right) _{\mu
_{1}},\left( A_{2}\right) _{\mu _{2}}\right] ^{A}.  \label{three2}
\end{align}%
The corresponding $L_{\infty }$ algebra is given by Eqs. (\ref{L1}) - (\ref%
{LE}) together with Eqs. (\ref{three1}) and (\ref{three2}). This reproduces
the algebraic formulation of 3D CS theory from Ref. \cite{Zwie}.

\section{$L_{\infty}$ formulation of FDA-based theories}

From now on, we will focus on gauge theories whose symmetries are not
described by Lie algebras but FDAs. The procedure to obtain the $L_{\infty }$
formulation of extended gauge theory is analogous to the one shown in the
previous case. However, the gauge subalgebra $L_{\infty }^{\mathrm{gauge}}$
will no longer be a Lie algebra and it cannot be trivially reduced into one
by writing the higher-degree forms in terms of one-forms as long as the
cocycle with which the algebra is defined is non-trivial. As we have seen,
the gauge transformations are given in Eqs. (\ref{d1}) and (\ref{d2}) by the
standard and extended versions covariant derivatives of a $0$-form and a $%
\left( p-1\right) $-form gauge parameters. Since the gauge variation of the
one-form $A^{A}$ is the same as with Lie algebras, the $L_{\infty }$
products obtained from its definition and the corresponding gauge algebra
are the same as in the previous section (see Eq. (\ref{gauge})). Thus, we will
focus on obtaining the products corresponding to the extended variations. As
before, we identify the gauge parameters and fields with the spaces $X_{0}$
and $X_{-1}$ respectively%
\begin{align}
\varepsilon & =\left( \varepsilon ^{A},\varepsilon ^{i}\right) \in X_{0}, \\
A& =\left( A^{A},A^{i}\right) \in X_{-1}.
\end{align}

\subsection{Gauge transformations}

The algebraic elements $\delta A$ carry both algebraic indices, each one
being a one-form and a $p$-form respectively. Therefore, the gauge variation
must be separated into its components $\delta A_{\mu}^{A}$ and $\delta
A_{\mu _{1}\cdots\mu_{p}}^{i}$. From Eq. (\ref{d2}) we can see that the
expression for $\delta A_{\mu_{1}\cdots\mu_{p}}^{i}$ in terms of $L_{\infty}$
products is truncated, resulting only in those terms that are powers of
degree zero, one and $p$ in the gauge fields%
\begin{equation}
\delta A_{\mu_{1}\cdots\mu_{p}}^{i}=\left[ \ell_{1}\left( \varepsilon
\right) \right] _{\mu_{1}\cdots\mu_{p}}^{i}+\left[ \ell_{2}\left(
\varepsilon,A\right) \right] _{\mu_{1}\cdots\mu_{p}}^{i}+\frac{\left(
-1\right) ^{\frac{p\left( p-1\right) }{2}}}{p!}\left[ \ell_{p+1}\left(
\varepsilon,A,\ldots,A\right) \right] _{\mu_{1}\cdots\mu_{p}}^{i}.
\end{equation}
Here we must point out an important difference with the previous section.
Each algebraic element on $X$ carries two differential forms of the
different degrees. Since the products between elements on $X$ are also
elements on $X$, they can be decomposed into its $A$- and $i$-components,
being those also differential forms of different degree. This is why some $%
L_{\infty}$ products carry a different number of coordinate indices
depending on which algebraic sector and subspace of $X$ they lie. For
simplicity, we introduce differential form products as follows 
\begin{equation}
\left[ \ell_{r}\left( x_{1},\ldots,x_{r}\right) \right] =\frac{1}{s!}\left[
\ell_{r}\left( x_{1},\ldots,x_{r}\right) \right] _{\mu _{1}\ldots\mu_{s}}%
\mathrm{d}x^{\mu_{1}}\wedge\cdots\wedge\mathrm{d}x^{\mu_{s}},   \label{diff}
\end{equation}
with $x_{1},\ldots,x_{n}\in X$ and $\ell_{r}\left( x_{1},\ldots,x_{r}\right) 
$ some non-vanishing $r$-linear product carrying $s$ antisymmetric
space-time indices. This allows to write the product between a large number
of elements without overloading of indices. One might then think that this
is an $L_{\infty}$ product between differential forms. However, this is not
the case; the product is still between algebraic elements that carry
differential forms of different degree and the result is being written in
terms of a basis for differential forms. From now on, we will write down the 
$L_{\infty}$ products in terms of differential forms. In the case of the
gauge variation of $A$ in the extended sector of the algebra this becomes%
\begin{equation}
\delta A^{i}=\left[ \ell_{1}\left( \varepsilon\right) \right] ^{i}+\left[
\ell_{2}\left( \varepsilon,A\right) \right] ^{i}+\frac{\left( -1\right) ^{%
\frac{p\left( p-1\right) }{2}}}{p!}\left[ \ell_{p+1}\left(
\varepsilon,A,\ldots,A\right) \right] ^{i},
\end{equation}
which leads to the following information about the $L_{\infty}$ products%
\begin{align}
\left[ \ell_{1}\left( \varepsilon\right) \right] ^{i} & =\mathrm{d}%
\varepsilon^{i},  \label{transf1} \\
\left[ \ell_{2}\left( \varepsilon,A\right) \right] ^{i} & =\left[
A,\varepsilon\right] ^{i}-\left[ \varepsilon,A\right] ^{i},   \label{transf2}
\\
\left[ \ell_{p+1}\left( \varepsilon,A^{p}\right) \right] ^{i} & =\left(
-1\right) ^{1+\frac{p\left( p-1\right) }{2}}\left[ \varepsilon ,A^{p}\right]
^{i}.   \label{transf3}
\end{align}
Every other product involving one element on $X_{0}$ and elements from $%
X_{-1}$ vanishes The next step is to obtain
the information concerning the gauge algebra.

\subsection{Gauge algebra}

As second step, we need to ensure the closure of the commutator of two gauge
transformations. By applying two consecutive transformations with parameters 
$\varepsilon_{1}=\left( \varepsilon_{1}^{A},\varepsilon_{1}^{i}\right) $ and 
$\varepsilon_{2}=\left( \varepsilon_{2}^{A},\varepsilon_{2}^{i}\right) $ and
taking the $i$-component we find%
\begin{align}
\left( \delta_{2}\delta_{1}-\delta_{1}\delta_{2}\right) A^{i} & =\left[ 
\mathrm{d}\varepsilon_{2},\varepsilon_{1}\right] ^{i}-\left[ \varepsilon
_{1},\mathrm{d}\varepsilon_{2}\right] ^{i}-\left[ \mathrm{d}\varepsilon
_{1},\varepsilon_{2}\right] ^{i}+\left[ \varepsilon_{2},\mathrm{d}%
\varepsilon_{1}\right] ^{i}  \notag \\
& +\left[ \varepsilon_{1},\left[ \varepsilon_{2},A\right] \right] ^{i}-\left[
\varepsilon_{2},\left[ \varepsilon_{1},A\right] \right] ^{i}+\left[ \left[
A,\varepsilon_{2}\right] ,\varepsilon_{1}\right] ^{i}-\left[ \left[
A,\varepsilon_{1}\right] ,\varepsilon_{2}\right] ^{i}+\left[ \varepsilon_{2},%
\left[ A,\varepsilon_{1}\right] \right] ^{i}-\left[ \varepsilon_{1},\left[
A,\varepsilon_{2}\right] \right] ^{i}  \notag \\
& +\frac{1}{p!}\left[ \varepsilon_{1},\left[ \varepsilon_{2},A^{p}\right] %
\right] ^{i}-\frac{1}{\left( p-1\right) !}\left[ \varepsilon_{1},\text{d}%
\varepsilon_{2},A^{p-1}\right] ^{i}-\frac{1}{\left( p-1\right) !}\left[
\varepsilon_{1},\left[ A,\varepsilon_{2}\right] ,A^{p-1}\right] ^{i}  \notag
\\
& -\frac{1}{p!}\left[ \varepsilon_{2},\left[ \varepsilon_{1},A^{p}\right] %
\right] ^{i}+\frac{1}{\left( p-1\right) !}\left[ \varepsilon_{2},\text{d}%
\varepsilon_{1},A^{p-1}\right] ^{i}+\frac{1}{\left( p-1\right) !}\left[
\varepsilon_{2},\left[ A,\varepsilon_{1}\right] ,A^{p-1}\right] ^{i}. 
\label{dd3}
\end{align}
On the other hand, from the generalized Jacobi identity it is possible to
prove the following relations%
\begin{align}
\left[ \varepsilon_{1},\left[ \varepsilon_{2},A\right] \right] ^{i}-\left[
\varepsilon_{2},\left[ \varepsilon_{1},A\right] \right] ^{i}-\left[ \left[
\varepsilon_{1},\varepsilon_{2}\right] ,A\right] ^{i} & =0,  \label{J1} \\
\left[ \varepsilon_{1},\left[ A,\varepsilon_{2}\right] \right] ^{i}+\left[ %
\left[ A,\varepsilon_{1}\right] ,\varepsilon_{2}\right] ^{i}-\left[ A,\left[
\varepsilon_{1},\varepsilon_{2}\right] \right] ^{i} & =0.   \label{J2}
\end{align}
Using Eqs. (\ref{J1}) and (\ref{J2}), the relation d$A^{A}=R^{A}-\frac{1}{2}%
\left[ A,A\right] ^{A}$ and integrating by parts, Eq. (\ref{dd3}) becomes 
\begin{equation}
\left( \delta_{2}\delta_{1}-\delta_{1}\delta_{2}\right) A^{i}=\delta
_{3}A^{i}-\frac{1}{\left( p-2\right) !}\left[ \varepsilon_{2},%
\varepsilon_{1},R,A^{p-2}\right] ^{i},   \label{trivR}
\end{equation}
where we introduce a third composite parameter $\varepsilon_{3}=\left(
\varepsilon_{3}^{A},\varepsilon_{3}^{i}\right) $ whose components depends on
the original parameters and gauge fields as follows%
\begin{align}
\varepsilon_{3}^{A} & =\left[ \varepsilon_{2},\varepsilon_{1}\right] ^{A},
\label{e3} \\
\varepsilon_{3}^{i} & =\left[ \varepsilon_{2},\varepsilon_{1}\right] ^{i}-%
\left[ \varepsilon_{1},\varepsilon_{2}\right] ^{i}+\frac{1}{\left(
p-1\right) !}\left[ \varepsilon_{2},\varepsilon_{1},A^{p-1}\right] ^{i}. 
\label{e3b}
\end{align}

Let us consider again two gauge transformations including both parameters~$%
\varepsilon^{A}$ and $\varepsilon^{i}$. In the $\ell$-picture, the
commutator between two gauge transformations is given by Eq. (\ref{ga0}) as
the sum of a trivial and non-trivial transformation. In order to extract the
relevant information, we compare Eqs. (\ref{ga0}) and (\ref{trivR}) to
truncate the expansion in terms of $L_{\infty}$ products. Then we can write $%
\delta_{\varepsilon_{3}}A$ in terms of a small set of products and split the
sum in a more convenient way, namely%
\begin{align}
\delta_{\varepsilon_{3}}A & =\ell_{1}\left( \varepsilon_{3}\right)
+\ell_{2}\left( \varepsilon_{3},A\right) +\frac{1}{p!}\left( -1\right) ^{%
\frac{p\left( p-1\right) }{2}}\ell_{p+1}\left( \varepsilon_{3},A^{p}\right) 
\notag \\
& =\left[ \delta_{\varepsilon_{3}}A\right] _{0}+\left[ \delta
_{\varepsilon_{3}}A\right] _{1}+\left[ \delta_{\varepsilon_{3}}A\right]
_{p-1}+\left[ \delta_{\varepsilon_{3}}A\right] _{p},   \label{de3}
\end{align}
where we denote $\left[ \delta_{\varepsilon_{3}}A\right] _{k}$ to the sum of
terms on $\delta_{\varepsilon_{3}}A$ with power $k$ in $A$. By plugging in
Eq. (\ref{e3}) and (\ref{e3b}) into Eq. (\ref{de3}) we obtain an explicit
expression for each term:%
\begin{align}
\left[ \delta_{\varepsilon_{3}}A\right] _{0} & =\ell_{1}\left( \ell
_{2}\left( \varepsilon_{1},\varepsilon_{2}\right) \right) ,  \label{var1} \\
\left[ \delta_{\varepsilon_{3}}A\right] _{1} & =\ell_{1}\left( \ell
_{3}\left( \varepsilon_{1},\varepsilon_{2},A\right) \right) +\ell _{2}\left(
\ell_{2}\left( \varepsilon_{1},\varepsilon_{2}\right) ,A\right) ,
\label{var2} \\
\left[ \delta_{\varepsilon_{3}}A\right] _{p-1} & =\frac{\left( -1\right) ^{%
\frac{\left( p-1\right) \left( p-2\right) }{2}}}{\left( p-1\right) !}%
\ell_{1}\left( \ell_{p+1}\left(
\varepsilon_{1},\varepsilon_{2},A^{p-1}\right) \right) +\frac{\left(
-1\right) ^{\frac{\left( p-2\right) \left( n-3\right) }{2}}}{\left(
p-2\right) !}\ell_{2}\left( \ell_{p}\left(
\varepsilon_{1},\varepsilon_{2},A^{p-2}\right) ,A\right) ,  \label{var3} \\
\left[ \delta_{\varepsilon_{3}}A\right] _{p} & =\frac{\left( -1\right) ^{%
\frac{p\left( p-1\right) }{2}}}{p!}\ell_{1}\left( \ell_{p+2}\left(
\varepsilon_{1},\varepsilon_{2},A^{p}\right) \right) +\frac{\left( -1\right)
^{\frac{\left( p-1\right) \left( p-2\right) }{2}}}{\left( p-1\right) !}%
\ell_{2}\left( \ell_{p+1}\left( \varepsilon_{1},\varepsilon
_{2},A^{p-1}\right) ,A\right)  \notag \\
& +\frac{\left( -1\right) ^{\frac{p\left( p-1\right) }{2}}}{p!}\ell
_{p+1}\left( \ell_{2}\left( \varepsilon_{1},\varepsilon_{2}\right)
,A^{p}\right) .   \label{var4}
\end{align}
From Eqs. (\ref{e3}) and (\ref{e3b}) it follows that the variation with
respect to $\varepsilon_{3}$ can be written in terms of $\varepsilon_{1}$
and $\varepsilon_{2}$ as%
\begin{align}
\delta_{3}A^{i} & =\mathrm{d}\left\{ \left[ \varepsilon_{2},\varepsilon _{1}%
\right] ^{i}-\left[ \varepsilon_{1},\varepsilon_{2}\right] ^{i}+\frac{1}{%
\left( p-1\right) !}\left[ \varepsilon_{2},\varepsilon _{1},A^{p-1}\right]
^{i}\right\} +\left[ A,\left\{ \left[ \varepsilon _{2},\varepsilon_{1}\right]
-\left[ \varepsilon_{1},\varepsilon_{2}\right] +\frac{1}{\left( p-1\right) !}%
\left[ \varepsilon_{2},\varepsilon _{1},A^{p-1}\right] \right\} \right] ^{i}
\notag \\
& -\left[ \left[ \varepsilon_{2},\varepsilon_{1}\right] ,A\right] ^{i}-\frac{%
1}{p!}\left[ \left[ \varepsilon_{2},\varepsilon_{1}\right] ,A^{p}\right]
^{i}.   \label{delta3}
\end{align}
We now compare Eq. (\ref{delta3}) with Eqs. (\ref{var1}) - (\ref{var4}) to
obtain four relations.

\emph{First relation (power }$0$\emph{\ in }$A$\emph{):}%
\begin{equation}
\left[ \ell_{1}\left( \ell_{2}\left( \varepsilon_{1},\varepsilon _{2}\right)
\right) \right] ^{i}=\mathrm{d}\left\{ \left[ \varepsilon
_{2},\varepsilon_{1}\right] ^{i}-\left[ \varepsilon_{1},\varepsilon _{2}%
\right] ^{i}\right\} .   \label{first}
\end{equation}

\emph{Second relation (power }$1$\emph{\ in }$A$\emph{):}%
\begin{equation}
\left[ \ell_{1}\left( \ell_{3}\left( \varepsilon_{1},\varepsilon
_{2},A\right) \right) +\ell_{2}\left( \ell_{2}\left( \varepsilon
_{1},\varepsilon_{2}\right) ,A\right) \right] ^{i}=\left[ A,\left( \left[
\varepsilon_{2},\varepsilon_{1}\right] -\left[ \varepsilon
_{1},\varepsilon_{2}\right] \right) \right] ^{i}-\left[ \left[
\varepsilon_{2},\varepsilon_{1}\right] ,A\right] ^{i}.   \label{sec}
\end{equation}

\emph{Third relation (power }$p-1$\emph{\ in }$A$\emph{):}%
\begin{align}
& \left( -1\right) ^{\frac{\left( p-1\right) \left( p-2\right) }{2}}\left[
\ell_{1}\left( \ell_{p+1}\left( \varepsilon_{1},\varepsilon
_{2},A^{p-1}\right) \right) \right] ^{i}+\left( -1\right) ^{\frac{\left(
p-2\right) \left( n-3\right) }{2}}\left( p-1\right) \left[ \ell _{2}\left(
\ell_{p}\left( \varepsilon_{1},\varepsilon_{2},A^{p-2}\right) ,A\right) %
\right] ^{i}  \notag \\
& =\text{d}\left[ \varepsilon_{2},\varepsilon_{1},A^{p-1}\right] ^{i}. 
\label{third}
\end{align}

\emph{Fourth relation (power }$p$\emph{\ in }$A$\emph{):}%
\begin{align}
& \left[ \ell_{1}\left( \ell_{p+2}\left( \varepsilon_{1},\varepsilon
_{2},A^{p}\right) \right) \right] ^{i}+p\left[ \ell_{2}\left( \ell
_{p+1}\left( \varepsilon_{1},\varepsilon_{2},A^{p-1}\right) ,A\right) \right]
^{i}+\left[ \ell_{p+1}\left( \ell_{2}\left( \varepsilon
_{1},\varepsilon_{2}\right) ,A^{p}\right) \right] ^{i}  \notag \\
& =\left( -1\right) ^{\frac{\left( p-1\right) \left( p-2\right) }{2}}p\left[
A,\left[ \varepsilon_{2},\varepsilon_{1},A^{p-1}\right] \right] ^{i}-\left(
-1\right) ^{\frac{\left( p-1\right) \left( p-2\right) }{2}}\left[ \left[
\varepsilon_{2},\varepsilon_{1}\right] ,A^{p}\right] ^{i}.   \label{fourth}
\end{align}

The products $\ell_{2}\left( \varepsilon_{1},\varepsilon_{2}\right) $, $%
\ell_{3}\left( \varepsilon_{1},\varepsilon_{2},A\right) $, $\ell_{p}\left(
\varepsilon_{1},\varepsilon_{2},A^{p-2}\right) $, $\ell_{p+1}\left(
\varepsilon_{1},\varepsilon_{2},A^{p-1}\right) $ and $\ell_{p+2}\left(
\varepsilon_{1},\varepsilon_{2},A^{p}\right) $ lie in $X_{0}$, and we can
therefore use the information obtained from the definition of gauge
transformations contained into Eqs. (\ref{transf1}) - (\ref{transf2}) into
Eqs. (\ref{first}) - (\ref{fourth}) to obtain explicit expressions for them.
Moreover, we have to compare the trivial transformation in the right-hand
side from Eq. (\ref{ga0}) with the last term on Eq. (\ref{trivR}). This
allows to write a product involving an element $R\in X_{-2}$. \emph{\ }%
\begin{equation}
\frac{\left( -1\right) ^{\frac{\left( p-2\right) \left( p-3\right) }{2}}}{%
\left( p-2\right) !}\ell_{p+1}\left( \varepsilon_{1},\varepsilon
_{2},R,A^{p-2}\right) =-\frac{1}{\left( p-2\right) !}\left[ \varepsilon
_{2},\varepsilon_{1},R,A^{p-2}\right] ^{i}.
\end{equation}

\emph{Summary:}

At this point we have found the relevant information contained in the gauge
transformations and the closure rule of the gauge algebra. That information
is codified into the following products:%
\begin{equation}
\left. 
\begin{array}{l}
\left[ \ell_{1}\left( \varepsilon_{1}\right) \right] ^{A}=\mathrm{d}%
\varepsilon_{1}^{A}, \\ 
\left[ \ell_{1}\left( \varepsilon_{1}\right) \right] ^{i}=\mathrm{d}%
\varepsilon_{1}^{i}, \\ 
\left[ \ell_{2}\left( \varepsilon_{1},A_{1}\right) \right] ^{A}=\left[
A_{1},\varepsilon_{1}\right] ^{A}, \\ 
\left[ \ell_{2}\left( \varepsilon_{1},A_{1}\right) \right] ^{i}=\left[
A_{1},\varepsilon_{1}\right] ^{i}-\left[ \varepsilon_{1},A_{1}\right] ^{i},
\\ 
\left[ \ell_{p+1}\left( \varepsilon_{1},A_{1},\ldots,A_{p}\right) \right]
^{i}=\left( -1\right) ^{1+\frac{p\left( p-1\right) }{2}}\left[
\varepsilon,A_{1},\ldots,A_{p}\right] ^{i},%
\end{array}
\right\} \text{ }%
\begin{array}{c}
\text{Gauge} \\ 
\text{transformations}%
\end{array}
\label{gt-fda}
\end{equation}%
\begin{equation}
\left. 
\begin{array}{l}
\left[ \ell_{2}\left( \varepsilon_{1},\varepsilon_{2}\right) \right] ^{A}=%
\left[ \varepsilon_{2},\varepsilon_{1}\right] ^{A}, \\ 
\left[ \ell_{2}\left( \varepsilon_{1},\varepsilon_{2}\right) \right] ^{i}=%
\left[ \varepsilon_{2},\varepsilon_{1}\right] ^{i}-\left[ \varepsilon_{1},%
\varepsilon_{2}\right] ^{i}, \\ 
\left[ \ell_{p+1}\left( \varepsilon_{1},\varepsilon_{2},A_{1},\ldots
,A_{p-1}\right) \right] ^{i}=\left( -1\right) ^{\frac{\left( p-1\right)
\left( p-2\right) }{2}}\left[ \varepsilon_{2},\varepsilon_{1},A_{1},%
\ldots,A_{p-1}\right] ^{i}, \\ 
\left[ \ell_{p+1}\left(
\varepsilon_{1},\varepsilon_{2},E,A_{1},\ldots,A_{p-2}\right) \right]
^{i}=\left( -1\right) ^{\frac{\left( p-2\right) \left( p-3\right) }{2}}\left[
\varepsilon_{1},\varepsilon _{2},E,A_{1},\ldots,A_{p-2}\right] ^{i},%
\end{array}
\right\} \text{ }%
\begin{array}{c}
\text{Gauge} \\ 
\multicolumn{1}{l}{\text{algebra}}%
\end{array}
\label{ga-fda}
\end{equation}
where $\varepsilon_{1},\varepsilon_{2}\in X_{0}$, $A_{1},\ldots,A_{p}\in
X_{-1}$ and $E\in X_{-2}$. Any other product involving elements on both
subspaces $X_{0}$ and $X_{-1}$ vanishes. These products will define an $%
L_{\infty}$ algebra and describe a consistent gauge theory if we include the
information coming from to the equations of motion. Note that different
theories with the same gauge symmetry will share the previously found
products. Starting from this point, we will consider two separate cases,
being the first one a flat FDA1 theory, in which the dynamics is governed by
the Maurer--Cartan equations, i.e., the zero-curvature conditions. The
second case to analyze will be the $q$-dimensional CS gauge theory invariant
under FDA1.

\subsection{Flat FDA1 theory}

In this case we consider zero-curvatures. This is not necessarily equivalent
to a flat space-time. The corresponding field equations are immediately
written as follows%
\begin{align}
\mathcal{F}^{A} & =\mathrm{d}A^{A}+\frac{1}{2}C_{BC}^{A}A^{B}A^{C},
\label{FA} \\
\mathcal{F}^{i} & =\mathrm{d}A^{i}+C_{Aj}^{i}A^{A}A^{j}+\frac{1}{\left(
p+1\right) !}C_{A_{1}\cdots A_{p+1}}^{i}A^{A_{1}}\cdots A^{A_{p+1}}. 
\label{Fi}
\end{align}
By comparing Eqs. (\ref{FA}) and (\ref{Fi}) with the general equation of
motion of Eq. (\ref{F1}) we can see that the expansion in terms of $%
L_{\infty }$ products gets truncated in different ways for both algebraic
sectors%
\begin{align}
\mathcal{F}^{A} & =\left[ \ell_{1}\left( A\right) \right] ^{A}-\frac {1}{2}%
\left[ \ell_{2}\left( A^{2}\right) \right] ^{A}, \\
\mathcal{F}^{i} & =\left[ \ell_{1}\left( A\right) \right] ^{i}-\frac {1}{2}%
\left[ \ell_{2}\left( A^{2}\right) \right] ^{i}+\frac{\left( -1\right) ^{%
\frac{p\left( p+1\right) }{2}}}{\left( p+1\right) !}\left[ \ell_{p+1}\left(
A^{p+1}\right) \right] ^{i}.
\end{align}
We can now obtain the information of the $L_{\infty}$ products acting on $%
X_{-1}$. For the Lie sector we get%
\begin{align}
\left[ \ell_{1}\left( A\right) \right] ^{A} & =\mathrm{d}A^{A}, \\
\left[ \ell_{2}\left( A^{2}\right) \right] ^{A} & =-C_{BC}^{A}A^{B}A^{C},
\end{align}
while for the extended sector we obtain one extra non-vanishing product,
carrying the information of the cocycle%
\begin{align}
\left[ \ell_{1}\left( A\right) \right] ^{i} & =\mathrm{d}A^{i}, \\
\left[ \ell_{2}\left( A^{2}\right) \right] ^{i} & =-2C_{Aj}^{i}A^{A}A^{j}, \\
\left[ \ell_{p+1}\left( A^{p+1}\right) \right] ^{i} & =\left( -1\right) ^{%
\frac{p\left( p+1\right) }{2}}C_{A_{1}\cdots A_{p+1}}^{i}A^{A_{1}}\cdots
A^{A_{p+1}}.
\end{align}
The dynamical sector for this theory is then summarized by the following
products%
\begin{equation}
\left. 
\begin{array}{l}
\left[ \ell_{1}\left( A\right) \right] ^{A}=\mathrm{d}A^{A}, \\ 
\left[ \ell_{1}\left( A\right) \right] ^{i}=\mathrm{d}A^{i}, \\ 
\left[ \ell_{2}\left( A_{1},A_{2}\right) \right]
^{A}=-C_{BC}^{A}A_{1}^{B}A_{2}^{C}, \\ 
\left[ \ell_{2}\left( A_{1},A_{2}\right) \right] ^{i}=-C_{Aj}^{i}\left(
A_{1}^{A}A_{2}^{j}+A_{2}^{A}A_{1}^{j}\right) , \\ 
\left[ \ell_{p+1}\left( A_{1},\ldots,A_{p+1}\right) \right] ^{i}=\left(
-1\right) ^{\frac{p\left( p+1\right) }{2}}C_{A_{1}\cdots
A_{p+1}}^{i}A_{1}^{A_{1}}\cdots A_{p+1}^{A_{p+1}},%
\end{array}
\right\} \text{ }%
\begin{array}{c}
\text{Flat FDA1} \\ 
\text{dynamical sector}%
\end{array}
\text{ }   \label{dflat}
\end{equation}%
\begin{equation}
\left. 
\begin{array}{l}
\left[ \ell_{2}\left( \varepsilon,E\right) \right] ^{A}=\left[ E,\varepsilon%
\right] ^{A}, \\ 
\left[ \ell_{2}\left( \varepsilon,E\right) \right] ^{i}=\left[ E,\varepsilon%
\right] ^{i}-\left[ \varepsilon,E\right] ^{i}, \\ 
\left[ \ell_{p+1}\left( \varepsilon,E,A_{1},\ldots,A_{p-1}\right) \right]
^{i}=\left( -1\right) ^{1+\frac{\left( p-1\right) \left( p-2\right) }{2}}%
\left[ \varepsilon,E,A_{1},\ldots,A_{p-1}\right] ^{i},%
\end{array}
\right\} \text{ }%
\begin{array}{c}
\text{Consistency} \\ 
\text{products}%
\end{array}
\label{c-fda}
\end{equation}
where $\varepsilon\in X_{0}$, $A,A_{1},\ldots,A_{p+1}\in X_{-1}$ and $E\in
X_{-2}$. As before, the consistency products involve elements on $X_{-2}$
and are obtained by plugging in the already found products into the $%
L_{\infty}$ identities. An explicit calculation can be found in Appendix
B.2. Eqs. (\ref{gt-fda}), (\ref{ga-fda}), (\ref{c-fda}) and (\ref{dflat})
define a consistent $L_{\infty}$ algebra encoding the information of the
flat FDA1 gauge theory. Note that this is consistent with the trivial gauge
transformation found in the gauge algebra from Eq. (\ref{trivR}). The
trivial gauge transformations vanish on-shell ensuring the closure of the
gauge subalgebra $L_{\infty}^{\mathrm{gauge}}$and thus the whole $L_{\infty}$
algebra.

\section{$L_{\infty}$ formulation of FDA1-CS theory}

Let us now consider extended CS theories. Since the gauge symmetry is
described, as in the previous section, by FDAs, it is not necessary to
reobtain most of the $L_{\infty}$ products but only those related with the
equations of motion. As before, it will be necessary to split the algebraic
elements into its $A$-, and $i$-components. Let us introduce the algebraic
elements $u=\left( u_{\mu}^{A},u_{\mu_{1}\cdots\mu_{p}}^{i}\right) \in X_{-1}
$ and $v=\left( v_{\mu_{1}\cdots\mu_{q-2}}^{A},v_{\mu_{1}\cdots
\mu_{q-p-1}}^{i}\right) \in X_{-2}$. Note that in this case we define the
indices for the elements on $X_{-2}$ in a different way that allows to
easily write the equations of motion in terms of them. We also define the
inner product between elements on these subspaces as%
\begin{equation}
\left\langle u,v\right\rangle _{L_{\infty}}=\int\mathrm{d}%
x^{2m-1}\varepsilon^{\mu_{1}\cdots\mu_{q-1}}\left(
g_{AB}u_{\mu_{1}}^{A}v_{\mu
_{2}\cdots\mu_{q-2}}^{B}+\varepsilon^{\mu_{1}\cdots\mu_{q-1}}g_{ij}u_{\mu
_{1}\cdots\mu_{p}}^{i}v_{\mu_{p+1}\cdots\mu_{q-1}}^{j}\right) ,
\end{equation}
where $g_{AB}$ and $g_{ij}$ the components of the rank-$2$ invariant tensor
defined on Eqs. (\ref{it1}) - (\ref{it3}). Components with mixed indices as $%
g_{Ai}$ are also allowed but they are not necessary in the definition of the
inner product. By setting $\bar{A}=0$ in Eq. (\ref{deltaS}) we can write the
variation of the action in terms of the inner product of the $L_{\infty}$
algebra as%
\begin{eqnarray}
\delta S &=&\left\langle \delta A,\mathcal{F}\right\rangle _{L_{\infty }} 
\notag \\
&=&\int \mathrm{d}x^{2m-1}\varepsilon ^{\mu _{1}\cdots \mu _{q-1}}\left( 
\frac{1}{\left( q-2\right) !}g_{AB}\delta A_{\mu _{1}}^{A}\mathcal{F}_{\mu
_{2}\cdots \mu _{q-2}}^{B}+\frac{1}{p!\left( q-p-1\right) !}\varepsilon
^{\mu _{1}\cdots \mu _{q-1}}g_{ij}\delta A_{\mu _{1}\cdots \mu _{p}}^{i}%
\mathcal{F}_{\mu _{p+1}\cdots \mu _{q-1}}^{j}\right)  \notag \\ \label{eom5}
\end{eqnarray}%
where the components of the algebraic element $\mathcal{F=}\left( \mathcal{F}%
_{\mu_{1}\cdots\mu_{q-3}}^{A},\mathcal{F}_{\mu_{1}\cdots\mu
_{q-p-1}}^{i}\right) \in X_{-2}$ can be written in terms of differential
forms as follows%
\begin{align}
\mathcal{F}^{A} & =\sum_{m,n}mg^{AA_{1}}g_{A_{1}A_{2}\cdots A_{m}i_{1}\cdots
i_{n}}R^{A_{2}}\cdots R^{A_{m}}R^{i_{1}}\cdots R^{i_{n}},  \label{F3} \\
\mathcal{F}^{i} & =\sum_{m,n}ng^{ii_{1}}g_{A_{1}\cdots A_{m}i_{1}i_{2}\cdots
i_{n}}R^{A_{1}}\cdots R^{A_{m}}R^{i_{2}}\cdots R^{i_{n}}.   \label{F4}
\end{align}
The indices $A$ and $i$ were raised in the original expressions (\ref{cseom1}%
) and (\ref{cseom2}) using the rank-$2$ invariant tensors $g^{AB}$ in the
case of the Lie sector and $g^{ij}$ for the extended one (see Appendix A).
This allows to get an explicit expression for $\mathcal{F}^{A}$ in terms of
the algebraic products.

\subsection{Lie sector}

Let us consider first the Lie sector of the equations of motion. From (\ref%
{F3}) and (\ref{F1}) we get an expression for $\mathcal{F}^{A}$, namely 
\begin{align}
\mathcal{F}^{A} & =\sum_{l=1}^{\infty}\frac{\left( -1\right) ^{\frac{l\left(
l-1\right) }{2}}}{l!}\left[ \ell_{l}\left( A^{l}\right) \right] ^{A}  \notag
\\
& =\sum_{m,n}mg_{\text{ \ }A_{2}\cdots A_{m}i_{1}\cdots
i_{n}}^{A}R^{A_{2}}\cdots R^{A_{m}}R^{i_{1}}\cdots R^{i_{n}}.
\end{align}
Moreover, by replacing the definition of curvatures (\ref{mu}) and (\ref{B})
into (\ref{F3}) we can explicitly write $\mathcal{F}^{A}$ in terms of the
gauge fields and their derivatives as%
\begin{align}
\mathcal{F}^{A} & =\sum_{m,n}\sum_{k=0}^{m-1}\sum_{r+s+t=n}\frac{1}{%
2^{k}\left( p+1\right) !^{t}}\frac{m!}{k!\left( m-k-1\right) !}\frac {n!}{%
r!s!t!}  \notag \\
& \times g_{\text{ \ }A_{1}\cdots A_{m-1}i_{1}\cdots i_{n}}^{A}\mathrm{d}%
A^{A_{1}}\cdots\mathrm{d}A^{m-k-1}\left[ A,A\right] ^{A_{m-k}}\cdots\left[
A,A\right] ^{A_{m-1}}  \notag \\
& \times\mathrm{d}A^{i_{1}}\cdots\mathrm{d}A^{i_{r}}\left[ A,A\right]
^{i_{r+1}}\cdots\left[ A,A\right] ^{i_{r+s}}\left[ A^{p+1}\right]
^{i_{s+r+1}}\cdots\left[ A^{p+1}\right] ^{i_{n}}.   \label{F6}
\end{align}
In order to isolate the contributions to $\mathcal{F}^{A}$ corresponding to
different $L_{\infty}$ products, we will separate the terms of the sum in
Eq. (\ref{F6}) that are powers of the same degree in the gauge fields, i.e., 
\begin{equation}
\mathcal{F}^{A}=\sum_{l=1}^{\infty}\left[ \mathcal{F}^{A}\right] _{l},
\end{equation}
with%
\begin{equation}
\left[ \mathcal{F}^{A}\right] _{l}=\frac{\left( -1\right) ^{\frac{l\left(
l-1\right) }{2}}}{l!}\left[ \ell_{l}\left( A^{l}\right) \right] ^{A}. 
\label{F7}
\end{equation}
Each term on the sum on the right-hand side of Eq. (\ref{F6}) is a power of
degree $m+n+k+s+pt-1$ in $A$. Thus, we can say that $\left[ \mathcal{F}^{A}%
\right] _{l}$ is equal to the sum of those terms verifying $m+n+k+s+pt-1=l$.
This allows us to write%
\begin{align}
\left[ \mathcal{F}^{A}\right] _{l} & =\sum_{m,n}\sum_{r+s+t=n}\frac {1}{%
2^{k_{st}}\left( p+1\right) !^{t}}\frac{m!}{k_{st}!\left( m-k_{st}-1\right) !%
}\frac{n!}{r!s!t!}  \notag \\
& \times g_{\text{ \ }A_{1}\cdots A_{m-1}i_{1}\cdots i_{n}}^{A}\mathrm{d}%
A^{A_{1}}\cdots\mathrm{d}A^{m-k_{st}-1}\left[ A,A\right] ^{A_{m-k_{st}}}%
\cdots\left[ A,A\right] ^{A_{m-1}}  \notag \\
& \times\mathrm{d}A^{i_{1}}\cdots\mathrm{d}A^{i_{r}}\left[ A,A\right]
^{i_{r+1}}\cdots\left[ A,A\right] ^{i_{r+s}}\left[ A^{p+1}\right]
^{i_{s+r+1}}\cdots\left[ A^{p+1}\right] ^{i_{n}},   \label{F8}
\end{align}
where $k_{st}=l+1-m-n-s-pt$. By comparing Eqs. (\ref{F7}) and (\ref{F8}) we
obtain the non vanishing $L_{\infty}$ products that describe the dynamics of
the theory for the Lie sector of the algebra 
\begin{align}
\left[ \ell_{l}\left( A^{l}\right) \right] ^{A} & =\left( -1\right) ^{\frac{%
l\left( l-1\right) }{2}}\sum_{m,n}\sum_{r+s+t=n}\frac{l!}{2^{k_{st}}\left(
p+1\right) !^{t}}\frac{m!}{k_{st}!\left( m-k_{st}-1\right) !}\frac{n!}{r!s!t!%
}  \notag \\
& \times g_{\text{ \ }A_{1}\cdots A_{m-1}i_{1}\cdots i_{n}}^{A}\mathrm{d}%
A^{A_{1}}\cdots\mathrm{d}A^{m-k_{st}-1}\left[ A,A\right] ^{A_{m-k_{st}}}%
\cdots\left[ A,A\right] ^{A_{m-1}}  \notag \\
& \times\mathrm{d}A^{i_{1}}\cdots\mathrm{d}A^{i_{r}}\left[ A,A\right]
^{i_{r+1}}\cdots\left[ A,A\right] ^{i_{r+s}}\left[ A^{p+1}\right]
^{i_{s+r+1}}\cdots\left[ A^{p+1}\right] ^{i_{n}}.   \label{dyn1}
\end{align}

\subsection{Extended sector}

For the equation of motion corresponding to the extended sector, we get a
similar expression. From Eq. (\ref{F4}) and the definition of curvatures, we
write the equation of motion in terms of the fields and their derivatives%
\begin{align}
\mathcal{F}^{i} & =\sum_{m,n}\sum_{k=0}^{m}\sum_{r+s+t=n-1}\frac{1}{%
2^{k}\left( p+1\right) !^{t}}\frac{m!}{k!\left( m-k\right) !}\frac {n!}{%
r!s!t!}g_{\text{ \ }i_{1}\cdots i_{n-1}A_{1}\cdots A_{m}}^{i}  \notag \\
& \times\mathrm{d}A^{A_{1}}\cdots\mathrm{d}A^{A_{m-k}}\left[ A,A\right]
^{A_{m-k+1}}\cdots\left[ A,A\right] ^{A_{m}}  \notag \\
& \times\mathrm{d}A^{i_{1}}\cdots\mathrm{d}A^{i_{r}}\left[ A,A\right]
^{i_{r+1}}\cdots\left[ A,A\right] ^{i_{r+s}}\left[ A^{p+1}\right]
^{i_{r+s+1}}\cdots\left[ A^{p+1}\right] ^{i_{n-1}}.
\end{align}
We now need to extract the part of the sum that has the same order on $A$.
Each term in the sum is a power of degree $m+n+k+s+pt-1$ in $A$. The terms
on $\mathcal{F}^{i}$ that are powers of degree $l$ are then given by

\begin{align}
\left[ \mathcal{F}^{i}\right] _{l} & =\sum_{m,n}\sum_{r+s+t=n-1}\frac {1}{2^{%
\bar{k}_{st}}\left( p+1\right) !^{t}}\frac{m!}{\bar{k}_{st}!\left( m-\bar{k}%
_{st}\right) !}\frac{n!}{r!s!t!}g_{\text{ \ }i_{1}\cdots i_{n-1}A_{1}\cdots
A_{m}}^{i}  \notag \\
& \times\mathrm{d}A^{A_{1}}\cdots\mathrm{d}A^{A_{m-\bar{k}_{st}}}\left[ A,A%
\right] ^{A_{m-\bar{k}_{st}+1}}\cdots\left[ A,A\right] ^{A_{m}}  \notag \\
& \times\mathrm{d}A^{i_{1}}\cdots\mathrm{d}A^{i_{r}}\left[ A,A\right]
^{i_{r+1}}\cdots\left[ A,A\right] ^{i_{r+s}}\left[ A^{p+1}\right]
^{i_{r+s+1}}\cdots\left[ A^{p+1}\right] ^{i_{n-1}},
\end{align}
where $\bar{k}_{st}=l+1-m-n-s-pt$. Since there are no symmetry or
antisymmetry rules for the different kind of indices on $g_{A_{1}\cdots
A_{m}i_{1}\cdots i_{n}}$, we have indistinctly denoted both sets of indices
in different order, i.e., $g_{A_{1}\cdots A_{m}i_{1}\cdots
i_{n}}=g_{i_{1}\cdots i_{n}A_{1}\cdots A_{m}}$.

By comparing with the general expression for $\mathcal{F}$ in Eq. (\ref{F1})
we get an expression for the products between gauge fields%
\begin{align}
\left[ \ell_{l}\left( A^{l}\right) \right] ^{i} & =\left( -1\right) ^{\frac{%
l\left( l-1\right) }{2}}\sum_{m,n}\sum_{r+s+t=n-1}\frac{l!}{2^{\bar{k}%
_{st}}\left( p+1\right) !^{t}}\frac{m!}{\bar{k}_{st}!\left( m-\bar{k}%
_{st}\right) !}\frac{n!}{r!s!t!}  \notag \\
& \times g_{\text{ \ }i_{1}\cdots i_{n-1}A_{1}\cdots A_{m}}^{i}\mathrm{d}%
A^{A_{1}}\cdots\mathrm{d}A^{A_{m-\bar{k}_{st}}}\left[ A,A\right] ^{A_{m-\bar{%
k}_{st}+1}}\cdots\left[ A,A\right] ^{A_{m}}  \notag \\
& \times\mathrm{d}A^{i_{1}}\cdots\mathrm{d}A^{i_{r}}\left[ A,A\right]
^{i_{r+1}}\cdots\left[ A,A\right] ^{i_{r+s}}\left[ A^{p+1}\right]
^{i_{r+s+1}}\cdots\left[ A^{p+1}\right] ^{i_{n-1}}.   \label{dyn2}
\end{align}

Eqs. (\ref{dyn1}) and (\ref{dyn2}) describe the complete dynamical sector of
the theory. As in the previous cases, for the products above to verify the $%
L_{\infty}$ identities, it is necessary to include some consistency products
involving elements on $X_{-2}$. Such products are given by%
\begin{equation}
\left. 
\begin{array}{l}
\left[ \mathcal{\ell}_{2}\left( \varepsilon,E\right) \right]
^{A}=C_{BC}^{A}E^{B}\varepsilon^{C}-g^{AB}g_{ij}C_{Bk}^{i}E^{j}%
\varepsilon^{k}, \\ 
\left[ \mathcal{\ell}_{p+1}\left( \varepsilon,E,A^{p-1}\right) \right]
^{A}=\left( -1\right) ^{1+\frac{\left( p-1\right) \left( p-2\right) }{2}%
}g^{AB_{1}}g_{ij}C_{B_{1}\cdots
B_{p+1}}^{i}\varepsilon^{B_{2}}A^{B_{3}}\cdots A^{B_{p+1}}E^{j}, \\ 
\left[ \mathcal{\ell}_{2}\left( \varepsilon,E\right) \right]
^{i}=g^{ij}g_{kl}C_{Bj}^{k}\varepsilon^{B}E^{l},%
\end{array}
\right\} \text{ }%
\begin{array}{c}
\text{Consistency} \\ 
\text{products}%
\end{array}
\label{c-csfda}
\end{equation}
with $\varepsilon\in X_{0}$, $A\in X_{-1}$ and $E\in X_{-2}$. An explicit
calculation for the consistency products can be found in Appendix B.3.

In summary, the complete set of products that define a consistent $L_{\infty}
$ algebra for FDA1-CS theory is given by Eqs. (\ref{gt-fda}) and (\ref%
{ga-fda}), describing the gauge symmetries, together with Eqs. (\ref{dyn1}),
(\ref{dyn2}) and (\ref{c-csfda}) describing the dynamical sector. For
simplicity, we do not explicitly write the product between arbitrary
elements on $X_{-1}$ as we did for the standard CS theory. Such expression
can be obtained from Eqs. (\ref{dyn1}) and (\ref{dyn2}) by considering
different algebraic elements in the argument of the products and including the corresponding symmetrization. Note
that, as it happens in the standard CS theory, there is a different number
of non-vanishing products in the dynamical sector, depending on the
dimensionality of the theory. Let us recall that the original CS action is $%
q-1$ dimensional and the values of $m$ and $n$ are the integer non-negative
solutions to the equation $2m+\left( p+1\right) n=q$. This implies that the
possible values of $l$ change case by case and they are restricted between $%
l_{\min}=n$ and $l_{\max}=2m-2$ for Eq. (\ref{dyn1}) (i.e., in the
Lie-sector) and $l_{\min}=n-1$ and $l_{\max}=2m$ for Eq. (\ref{dyn2}) (i.e.,
in the extended sector). This completes the algebraic products of the $%
L_{\infty}$ formulation of FDA1-CS theory. Contrary to the standard case,
the gauge subalgebra is not a Lie algebra but also an $L_{\infty}$ algebra.

\subsection{Five-dimensional theory}

As an example, let us consider a five-dimensional CS theory for an FDA1 with 
$p=3$ in absence of a cocycle$\,$. The corresponding FDA is given by the
following Maurer--Cartan equations%
\begin{align}
\mathrm{d}A^{A}+\frac{1}{2}C_{BC}^{A}A^{B}A^{C} & =R^{A}=0, \\
\mathrm{d}A^{i}+C_{Aj}^{i}A^{A}A^{j} & =R^{i}=0.
\end{align}
This allows to formulate a gauge theory with non-trivial coupling between a
one-form and a three-form whose gauge-invariant action is given by%
\begin{equation}
S_{5}\left[ A\right] =\int_{M_{5}}\int_{0}^{1}\mathrm{d}t\left(
3g_{A_{1}A_{2}A_{3}}A^{A_{1}}R_{t}^{A_{2}}R_{t}^{A_{3}}+g_{A_{1}i_{1}}A^{A_{1}}R_{t}^{i_{1}}+g_{A_{1}i_{1}}R_{t}^{A_{1}}A^{i_{1}}\right) , 
\label{S5}
\end{equation}
where $A_{t}=tA$, and being $R_{t}$ its corresponding field strength. An
example of this action for a particular bosonic FDA can be found in Ref. 
\cite{CSFDA}.

The $L_{\infty}$ products containing the information of the symmetries and
interacting theory are then given by%
\begin{equation}
\begin{tabular}{l}
$%
\begin{tabular}{l}
$\left[ \ell_{1}\left( \varepsilon\right) \right] _{\mu}^{A}=\partial
_{\mu}\varepsilon^{A},$ \\ 
$\left[ \ell_{1}\left( \varepsilon\right) \right] _{\mu}^{i}=\partial
_{\mu}\varepsilon^{i},$ \\ 
$\left[ \ell_{2}\left( \varepsilon,A\right) \right] _{\mu}^{A}=C_{BC}^{A}A_{%
\mu}^{B}\varepsilon^{C},$ \\ 
$\left[ \ell_{2}\left( \varepsilon,A\right) \right] _{\mu\nu%
\rho}^{i}=C_{Aj}^{i}\left(
3A_{[\mu}^{A}\varepsilon_{\nu\rho]}^{j}-\varepsilon
^{A}A_{\mu\nu\rho}^{j}\right) ,$ \\ 
\ 
\end{tabular}
\ $ \\ 
\multicolumn{1}{c}{$\text{Gauge transformations}$}%
\end{tabular}
\ \text{ \ \ }%
\begin{tabular}{l}
$%
\begin{tabular}{l}
$\left[ \ell_{2}\left( \varepsilon_{1},\varepsilon_{2}\right) \right]
^{A}=C_{BC}^{A}\varepsilon_{2}^{B}\varepsilon_{1}^{C},$ \\ 
$\left[ \ell_{2}\left( \varepsilon_{1},\varepsilon_{2}\right) \right]
^{A}=C_{BC}^{A}\varepsilon_{2}^{B}\varepsilon_{1}^{C},$ \\ 
$\left[ \ell_{2}\left( \varepsilon_{1},\varepsilon_{2}\right) \right]
_{\mu\nu}^{i}=C_{Aj}^{i}\left( \varepsilon_{2}^{A}\left( \varepsilon
_{1}\right) _{\mu\nu}^{j}-\varepsilon_{1}^{A}\left( \varepsilon_{2}\right)
_{\mu\nu}^{j}\right) .$ \\ 
\ \ 
\end{tabular}
\ $ \\ 
\multicolumn{1}{c}{Gauge algebra}%
\end{tabular}%
\end{equation}
The allowed values for $\left( m,n\right) $ are $\left( 3,0\right) $ and $%
\left( 1,1\right) $ and therefore, from Eqs. (\ref{dyn1}) and (\ref{dyn2})
we see that there are three non-vanishing products for the dynamical sector
in the Lie sector of the algebra and one product for the extended one%
\begin{equation*}
\left. 
\begin{array}{l}
\left[ \ell_{4}\left( A_{1},A_{2},A_{3},A_{4}\right) \right] _{\mu\nu
\rho\sigma}^{A}=3\times3!%
\times4!g^{AD}g_{DBC}C_{B_{1}B_{2}}^{B}C_{C_{1}C_{2}}^{C}\left(
A_{\{1}\right) _{[\mu}^{B_{1}}\left( A_{2}\right) _{\nu }^{B_{2}}\left(
A_{3}\right) _{\rho}^{C_{1}}\left( A_{4\}}\right) _{\sigma]}^{C_{2}}, \\ 
\left[ \ell_{3}\left( A_{1},A_{2},A_{3}\right) \right] _{\mu\nu\rho\sigma
}^{A}=-3\times3!\times4!g^{AF}g_{FBC}C_{DE}^{B}\left( A_{\{1}\right) _{[\mu
}^{D}\left( A_{2}\right) _{\nu}^{E}\partial_{\rho}\left( A_{3\}}\right)
_{\sigma]}^{C}, \\ 
\left[ \ell_{2}\left( A_{1},A_{2}\right) \right] _{\mu\nu\rho\sigma}^{A}=-3!%
\times4!g^{AD}g_{DBC}\partial_{\lbrack\mu}\left( A_{\{1}\right)
_{\nu}^{B}\partial_{\rho}\left( A\right) _{\sigma]}^{C}-8g_{\text{ \ }%
i}^{A}C_{Bj}^{i}\left( A_{\{1}\right) _{[\mu}^{B}\left( A_{2\}}\right)
_{\nu\rho\sigma]}^{j}, \\ 
\left[ \ell_{2}\left( A_{1},A_{2}\right) \right] _{\mu%
\nu}^{i}=-2g^{ij}g_{Aj}C_{BC}^{A}\left( A_{1}\right) _{\{\mu}^{B}\left(
A_{2}\right) _{\nu\}}^{C},%
\end{array}
\right\} \ \text{ }%
\begin{array}{c}
\text{Dynamical} \\ 
\text{sector}%
\end{array}
\end{equation*}%
\begin{equation}
\left. 
\begin{array}{l}
\left[ \mathcal{\ell}_{2}\left( \varepsilon,E\right) \right] _{\mu\nu
\rho\sigma}^{A}=C_{BC}^{A}E_{\mu\nu\rho\sigma}^{B}%
\varepsilon^{C}-3!g^{AB}g_{ij}C_{Bk}^{i}E_{[\mu\nu}^{j}\varepsilon_{\rho%
\sigma]}^{k}, \\ 
\left[ \mathcal{\ell}_{2}\left( \varepsilon,E\right) \right]
_{\mu\nu}^{i}=2g^{ij}g_{kl}C_{Bj}^{k}\varepsilon^{B}E_{\mu\nu}^{l}.%
\end{array}
\right\} \text{ \ }%
\begin{array}{c}
\text{Consistency} \\ 
\text{products}%
\end{array}%
\end{equation}
A particular case of this theory is the so-called BF theory, in which only
the third term in the integral at the right side of Eq. (\ref{S5}) is not
vanishing.

\section{Concluding Remarks}

We have formulated the standard CS theory and two cases of FDA-based
theories in terms of $L_{\infty}$ algebras. In the standard CS case, we
wrote down the products between algebraic elements. Such products satisfy
the graded symmetry rule from Eq. (\ref{symml}). In later cases, for
FDA-based theories, the corresponding products were written in terms of
differential forms for simplicity. It is important to note that the
algebraic products satisfy the mentioned graded symmetry from Eq. (\ref%
{symml}) if we write them in terms of the gauge fields without using
differential forms. By writing the products in terms of differential forms,
we add a new algebraic structure, and therefore all the products found
starting from Section 5 satisfy a modified symmetry rule. The same
difference is present in the formulation of standard gauge theories, where
the Lie product is antisymmetric but, when dealing with differential forms,
it becomes symmetric or antisymmetric depending on their differential
degrees. The reason for this choice in the latter cases is that the
algebraic products take on a particularly simple form due to the natural
presence of higher-degree forms as gauge fields in FDA-based theories. As we can see in Eq. (\ref%
{diff}), the $L_{\infty}$ products, satisfying the graded-symmetry rule (\ref%
{symml}) can be easily found by removing the dependence on the basis of
differential forms and including the corresponding antisymmetrization. This
leads to larger expressions that we explicitly write for the
five-dimensional case in the example at the end of Section 6.

An FDA-based gauge theory can naturally involve higher-degree differential
forms as gauge fields due to the presence of higher-degree products in the dynamical sector.
However, an important issue must be noticed. Higher-degree components of the
field strength do not transform covariantly. This was pointed out as a
general feature of $L_{\infty}$ gauge theories involving higher-degree forms
in Ref. \cite{bonezzi}. As a consequence, the gauge algebra is closed only
on-shell, as shown in Eq. (\ref{ga0}) due to the presence of a trivial gauge
transformation that vanishes by imposing the equations of motion. The
definition of gauge transformations imposes then a constraint on the action.
In the case of FDA1, the gauge algebra closes on-shell only if the equations
of motions imply that the two-form curvature $R^{A}$ vanishes (there are no
constraints on the extended curvature $R^{i}$). Otherwise, such gauge transformations
do not close, and therefore, the $L_{\infty}$ algebra that describes the whole theory
has a missing definition in its gauge subalgebra. The generalized CS action remains
invariant under the FDA1 gauge transformations but it is not an entirely well-defined gauge theory.
The $L_{\infty}$ formulation of FDA1-CS theory is therefore valid, only for cases
in which the equations of motion are not inconsistent with the closure of
the gauge algebra. Simple examples of this can be found by choosing
symmetries described by trivial extensions of Lie algebras, i.e., when the
extended Maurer--Cartan equation does not include a non-trivial cocycle.
Moreover, if the theory is three-dimensional or the FDA invariant tensors do
not present mixed indices corresponding to both algebraic sectors (as $g_{Ai}
$), a non-trivial FDA with closed gauge algebra is also possible. In
contrast, the gauge algebra always close on-shell for the flat FDA theory
(this is also true for any other theory verifying $R^{A}=0$ on-shell). Since
the trivial gauge transformations on Eq. (\ref{t-g}) appear when dealing
with products of three or more elements, this is not an issue for algebras
with bilinear products either, as is the case of Lie algebras or in the
five-dimensional example of Section 6.

\section*{Acknowledgements}

The autor would like to thank Laura Adrianopoli, Fabrizio Cordonier-Tello,
Nicol\'{a}s Gonz\'{a}lez, Dieter L\"{u}st and Mario Trigiante for
enlightening discussions. This research was partially funded by the
bilateral DAAD-CONICYT grant 62160015. The author acknowledges the support
from the Max-Planck-Society.

\appendix

\section{Notation}

In this appendix we will introduce a compact notation for FDAs, similar to
the one usually used for brackets and invariant tensors of Lie algebras. Let 
$x\in\bar{X}$ be an algebraic element for certain FDA as it was introduced
in Section 3. This means that we consider a collection of differential forms 
\begin{equation}
x=\left( x^{A\left( 1\right) },\ldots,x^{A\left( N\right) }\right) ,
\end{equation}
each one belonging to a certain subspace of the graded space $\bar{X}\,$\ ($%
\deg_{\bar{X}}x^{A\left( p\right) }=p$). We introduce an FDA-degree that
will be useful to identity the differential-form degree of each component.
Such degree corresponds to the differential form degree of its first
component, i.e., if $x^{A\left( 1\right) }$ is a $r$-form, we say that $%
\deg_{\text{FDA}}x=r$. Therefore, we can say that each component $x^{A\left(
p\right) }$ is given by a $\left( r+p-1\right) $-form. The FDA-degree is
particularly useful in the case of the FDA from Eqs. (\ref{mu}) and \ref{B})
where $X$ is reduced to two subspaces and $x$ to the following pair 
\begin{equation}
x=\underset{r\text{-form}}{\left( x^{A\left( 1\right) }\right. },\underset{%
\left( r+p-1\right) \text{-form}}{\left. x^{A\left( p\right) }\right) }\text{
}.
\end{equation}
For convenience, will denote $x^{A\left( p\right) }$ as an $\bar{r}$-form
with $\bar{r}=r+p-1$.

We now introduce a product between algebraic elements in terms of the
brackets from Eq. (\ref{T}), which in this particular FDA case, is reduced
to a bilinear bracket and a $\left( p+1\right) $-linear one. Let us consider
a set of algebraic elements $B_{1},\ldots B_{p+1}$ on FDA1, each one of
FDA-degree $b_{1},\ldots,b_{p+1}$. In terms of the structure constants of
the FDA1, we define:

\begin{enumerate}
\item A bilinear product $\left[ B_{1},B_{2}\right] $ such that 
\begin{align}
\left[ B_{1},B_{2}\right] ^{A} & =C_{BC}^{A}B_{1}^{B}B_{2}^{C}, \\
\left[ B_{1},B_{2}\right] ^{i} & =C_{Bj}^{i}B_{1}^{B}B_{2}^{j}.
\end{align}

\item A $\left( p+1\right) $-linear product $\left[ B_{1},\ldots ,B_{p+1}%
\right] $ such that 
\begin{align}
\left[ B_{1},\ldots,B_{p+1}\right] ^{A} & =0, \\
\left[ B_{1},\ldots,B_{p+1}\right] ^{i} & =C_{A_{1}\cdots
A_{p+1}1}^{i}B_{1}^{A_{1}}\cdots B_{p+1}^{A_{p+1}}.
\end{align}
\end{enumerate}

Let us now consider a two sets of algebraic elements $B_{1},\ldots B_{m}$
and $E_{1},...,E_{n}$, each one of FDA-degree $b_{1},\ldots,b_{m}$ and $%
e_{1},\ldots e_{n}$ respectively. We introduce a compact notation for the
contraction of their algebraic components with the FDA1 invariant tensor 
\begin{equation}
\left\langle B_{1},\ldots,B_{m};E_{1},\ldots,E_{n}\right\rangle
=g_{A_{1}\cdots A_{m}i_{1}\cdots i_{n}}B_{1}^{A_{1}}\cdots
B_{m}^{A_{m}}E_{1}^{i_{1}}\cdots E_{n}^{i_{n}}.
\end{equation}
This bracket is equivalent to the symmetrized trace notation for Lie
algebras for $n=0$. In general, it separates the algebraic sectors before
and after the semicolon, being the first ones evaluated in the Lie sector
and the latter in the extended sector. In this notation, the following
properties are fulfilled%
\begin{align}
\left\langle \ldots,B_{r},B_{r+1},\ldots;E_{1},\ldots,E_{n}\right\rangle &
=\left( -1\right) ^{b_{r}b_{r+1}}\left\langle
\ldots,B_{r+1},B_{r},\ldots;E_{1},\ldots,E_{n}\right\rangle , \\
\left\langle B_{1},\ldots,B_{m};\ldots,E_{s},E_{s+1},\ldots\right\rangle &
=\left( -1\right) ^{\bar{e}_{s}\bar{e}_{s+1}+p+1}\left\langle
B_{1},\ldots,B_{m};\ldots,E_{s+1},E_{s},\ldots\right\rangle .
\end{align}

The invariant properties of $g_{A_{1}\cdots A_{m}i_{1}\cdots i_{n}}$ provide
us with notion of covariance and contravariance. Given an algebraic element $%
B=\left( B^{A},B^{i}\right) $ with FDA-degree $b$, we define the
contravariant duals of its components as $B_{A}=g_{AB}B^{B}$ and $%
B_{i}=g_{ij}B^{i}$. Note that, although the components with mixed indices $%
g_{Ai}$ are in general non-vanishing, we do not include them into the
definition in order to not change the differential form degree of the
components of $B$. In this way, $B_{A}$ and $B_{i}$ are still a $b$-form and 
$\bar{b}$-form respectively. In the same way, we define the inverse
components $g^{AB}$ and $g^{ij}$ through the following relations%
\begin{equation}
g_{AB}=g_{AC}g_{BD}g^{CD},\ \ \ \ g_{ij}=g_{ik}g_{jl}g^{kl}.
\end{equation}
In the standard case, $g_{AB}$ is reduced to the Cartan--Killing metric for
Lie algebras. Note that this is not a rigorous definition of a generalized
Cartan--Killing metric for FDA1 but a notation that will be useful when
writing products without introducing ambiguity.

\section{Consistency products on $X_{-2}$}

In this appendix, we will obtain the missing products in the $L_{\infty}$
formulation of CS theory and FDA-based gauge theories. Such products do not
contain information coming directly from the gauge transformations, gauge
algebra or equations of motion but must be non-vanishing for consistency.
They act on $X_{-2}$ and can be found by replacing the already found
products into the $L_{\infty}$ identities.

A simple procedure to find them is to use the definition Eq. (\ref{F1}) and
take its gauge variation: 
\begin{equation}
\delta\mathcal{F}=\sum_{k=1}^{\infty}\sum_{r=0}^{\infty}\frac{\left(
-1\right) ^{\frac{k\left( k-1\right) +r\left( r-1\right) }{2}}}{k!r!}%
\ell_{k}\left( \ell_{r+1}\left( \varepsilon,A^{r}\right) ,A^{n-1}\right) .
\end{equation}
Using the $L_{\infty}$ identities (\ref{Lid}) this can be written as%
\begin{equation}
\delta\mathcal{F=}\sum_{k=0}^{\infty}\frac{\left( -1\right) ^{\frac{k\left(
k-1\right) }{2}}}{k!}\mathcal{\ell}_{k+2}\left( \varepsilon,\mathcal{F}%
,A^{k}\right) .   \label{df}
\end{equation}
We can now directly compare this expression with the variation of $\mathcal{F%
}$ obtained case by case that contains the information about the products of 
$X_{-1}$. Let us treat this separately.

\subsection{Products in CS theory}

In this case, the algebraic element $\mathcal{F}\in X_{-2}$ is given by%
\begin{equation}
\mathcal{F}_{\nu}^{A}=\varepsilon_{\nu}^{\text{ \ }\mu_{1}\cdots%
\mu_{2m-2}}g_{\text{ \ }B_{1}\cdots
B_{m-1}}^{A}R_{\mu_{1}\mu_{2}}^{B_{1}}\cdots
R_{\mu_{2m-3}\mu_{2m-2}}^{B_{m-1}}.
\end{equation}
By plugging in the well-known relation $\delta
R_{\mu\nu}^{A}=C_{BC}^{A}R_{\mu \nu}^{B}\varepsilon^{C}$ we find%
\begin{align}
\delta\mathcal{F}_{\nu}^{A} & =\left( m-1\right) \varepsilon_{\nu}^{\text{ \ 
}\mu_{1}\cdots\mu_{2m-2}}g_{\text{ \ }B_{1}\cdots
B_{m-1}}^{A}C_{BC}^{B_{1}}R_{\mu_{1}\mu_{2}}^{B},\varepsilon^{C}R_{\mu_{2}%
\mu_{4}}^{B_{2}}\cdots R_{\mu_{2m-3}\mu_{2m-2}}^{B_{m-1}}  \notag \\
& =\left( m-1\right) \varepsilon_{\nu}^{\text{ \ }\mu_{1}\cdots\mu_{2m-2}}g_{%
\text{ \ }B_{1}\cdots
B_{m-1}}^{A}C_{BC}^{B_{1}}R_{\mu_{1}\mu_{2}}^{B},\varepsilon^{C}R_{\mu_{2}%
\mu_{4}}^{B_{2}}\cdots R_{\mu_{2m-3}\mu_{2m-2}}^{B_{m-1}}.
\end{align}
Using the definition of invariant tensor for Lie algebras (such definition
can be obtained by setting $n=0$ on Eq. (\ref{it1})) we find%
\begin{align}
\delta\mathcal{F}_{\nu}^{A} & =\varepsilon_{\nu}^{\text{ \ }\mu_{1}\cdots
\mu_{2m-2}}g_{\text{ \ }B_{1}\cdots B_{m-1}}^{B}C_{BC}^{A}\varepsilon
^{C}R_{\mu_{1}\mu_{2}}^{B_{1}}\cdots R_{\mu_{2m-3}\mu_{2m-2}}^{B_{m-1}} 
\notag \\
& =C_{BC}^{A}\mathcal{F}_{\nu}^{B}\varepsilon^{C}.   \label{dfcs}
\end{align}
This shows that, in the case of the standard CS theory, the
equation-of-motion term $\mathcal{F}_{\nu}^{A}$ inherits the transformation
law of the $2$-form curvature. By directly comparing Eqs. (\ref{df}) and (%
\ref{dfcs}) we obtain the following product involving elements on $X_{0}$
and $X_{-2}$:%
\begin{equation}
\left[ \mathcal{\ell}_{2}\left( \varepsilon,\mathcal{F}\right) \right]
^{A}=C_{BC}^{A}\mathcal{F}^{B}\varepsilon^{C}.
\end{equation}

\subsection{Products in FDA-flat theory}

In this case, the equations of motions are equivalent to the Maurer--Cartan
equations for FDA1, i.e., $\left( \mathcal{F}^{A},\mathcal{F}^{i}\right)
=\left( R^{A},R^{i}\right) $. The gauge variation of $R$ under a
transformation with parameter $\varepsilon=\left(
\varepsilon^{A},\varepsilon^{i}\right) $ is then given by%
\begin{align}
\delta R^{A} & =C_{BC}^{A}R^{B}\varepsilon^{C},  \label{deltaf1} \\
\delta R^{i} & =C_{Aj}^{i}R^{A}\varepsilon^{j}-C_{Aj}^{i}\varepsilon
^{A}R^{j}-\frac{1}{\left( p-1\right) !}C_{A_{1}\cdots
A_{p+1}}^{i}\varepsilon^{A_{1}}R^{A_{2}}A^{A_{3}}\cdots A^{A_{p+1}}. 
\label{deltaf2}
\end{align}
By comparing Eqs. (\ref{deltaf1}), (\ref{deltaf2}) and (\ref{df}) we
immediately find the following brackets (as mentioned in Section 5, we write
the products in terms of differential forms for FDA-based theories)%
\begin{align}
\left[ \mathcal{\ell}_{2}\left( \varepsilon,\mathcal{F}\right) \right] ^{A}
& =C_{BC}^{A}\mathcal{F}^{B}\varepsilon^{C}, \\
\left[ \mathcal{\ell}_{2}\left( \varepsilon,\mathcal{F}\right) \right] ^{i}
& =C_{Aj}^{i}\left( \mathcal{F}^{A}\varepsilon^{j}-\varepsilon ^{A}\mathcal{F%
}^{j}\right) , \\
\left[ \mathcal{\ell}_{p+1}\left( \varepsilon,\mathcal{F},A^{p-1}\right) %
\right] ^{i} & =\left( -1\right) ^{1+\frac{\left( p-1\right) \left(
p-2\right) }{2}}C_{A_{1}\cdots A_{p+1}}^{i}\varepsilon^{A_{1}}\mathcal{F}%
^{A_{2}}A^{A_{3}}\cdots A^{A_{p+1}}.
\end{align}

\subsection{Products in FDA-CS theory}

Unlike the previous ones, the algebraic element $\mathcal{F}$ does not have
the same transformation law as the field strength $R$, leading to more
complicated expressions for the consistency products.

\subsubsection{Lie sector}

From the variation of the curvatures on Eqs. (\ref{deltaf1}), (\ref{deltaf2}%
) and the definition on Eq. (\ref{F3}) we obtain the gauge variation of $%
\mathcal{F}^{A}$:%
\begin{align}
\delta\mathcal{F}^{A} & =\sum_{m,n}mg^{AA_{1}}g_{A_{1}\cdots
A_{m}i_{1}\cdots i_{n}}\left( \left( m-1\right)
C_{BC}^{A_{2}}R^{B}\varepsilon^{C}R^{A_{3}}\cdots R^{A_{m}}R^{i_{1}}\cdots
R^{i_{n}}\right.  \notag \\
& +nR^{A_{2}}\cdots
R^{A_{m}}C_{Aj}^{i_{1}}R^{A}\varepsilon^{j}R^{i_{2}}\cdots
R^{i_{n}}-nR^{A_{2}}\cdots
R^{A_{m}}C_{Aj}^{i_{1}}\varepsilon^{A}R^{j}R^{i_{2}}\cdots R^{i_{n}}  \notag
\\
& \left. -\frac{n}{\left( p-1\right) !}R^{A_{2}}\cdots
R^{A_{m}}C_{A_{1}\cdots
A_{p+1}}^{i_{1}}\varepsilon^{A_{1}}R^{A_{2}}A^{A_{3}}\cdots
A^{A_{p+1}}R^{i_{2}}\cdots R^{i_{n}}\right) .   \label{dfa}
\end{align}
Using the invariant tensor conditions (\ref{it1}) - (\ref{it3}), it is
possible to prove the following relations:%
\begin{align}
0 & =g_{A_{1}\cdots A_{m}i_{1}\cdots i_{n}}\left( \left( m-1\right)
C_{BC}^{A_{2}}R^{B}\varepsilon^{C}R^{A_{3}}\cdots R^{A_{m}};R^{i_{1}}\cdots
R^{i_{n}}-nR^{A_{2}}\cdots
R^{A_{m}}C_{Bj}^{i_{1}}\varepsilon^{B}R^{j}R^{i_{2}}\cdots R^{i_{n}}\right) 
\notag \\
& -g_{AA_{2}\cdots A_{m}i_{1}\cdots i_{n}}C_{BA_{1}}^{A}\varepsilon
^{B}R^{A_{2}}\cdots R^{A_{m}}R^{i_{1}}\cdots R^{i_{n}},   \label{id1}
\end{align}%
\begin{align}
0 & =mg_{A_{1}\cdots A_{m}i_{1}\cdots i_{n}}f^{A_{1}}R^{A_{2}}\cdots
R^{A_{m}};C_{B_{1}\cdots
B_{p+1}}^{i_{1}}R^{B_{1}}\varepsilon^{B_{2}}A^{B_{3}}\cdots
A^{B_{p+1}}R^{i_{2}}\cdots R^{i_{n}}  \notag \\
& +g_{AA_{2}\cdots A_{m}i_{1}\cdots i_{n}}R^{A}R^{A_{2}}\cdots
R^{A_{m}}C_{A_{1}B_{2}\cdots
B_{p+1}}^{i_{1}}\varepsilon^{B_{2}}A^{B_{3}}\cdots
A^{B_{p+1}}R^{i_{1}}\cdots R^{i_{n}},   \label{id2}
\end{align}%
\begin{align}
0 & =mg_{A_{1}\cdots A_{m}i_{1}\cdots i_{n}}R^{A_{2}}\cdots
R^{A_{m}}C_{Bj}^{i_{1}}R^{B}\varepsilon^{j}R^{i_{2}}\cdots R^{i_{n}}  \notag
\\
& +g_{AA_{2}\cdots A_{m}i_{1}\cdots i_{n}}R^{A}R^{A_{2}}\cdots
R^{A_{m}}C_{A_{1}j}^{i_{1}}\varepsilon^{j}R^{i_{2}}\cdots R^{i_{n}}. 
\label{id3}
\end{align}
We replace Eqs. (\ref{id1}) - (\ref{id3}) into Eq. (\ref{dfa}) and write the
variation of $\mathcal{F}^{A}$ in terms of $\mathcal{F}^{A}$ and $\mathcal{F}%
^{i}$: 
\begin{equation}
\delta\mathcal{F}^{A}=C_{BC}^{A}\mathcal{F}^{B}%
\varepsilon^{C}-g^{AB}g_{ik}C_{Bj}^{i}\mathcal{F}^{k}\varepsilon^{j}-\frac{1%
}{\left( p-1\right) !}g^{AB_{1}}g_{ik}C_{B_{1}B_{2}\cdots
B_{p+1}}^{i}\varepsilon^{B_{2}}A^{B_{3}}\cdots A^{B_{p+1}}\mathcal{F}^{k}. 
\label{deltafa}
\end{equation}
By comparing Eqs. (\ref{deltafa}) and (\ref{df}) we obtain the following
products:%
\begin{align}
\left[ \mathcal{\ell}_{2}\left( \varepsilon,\mathcal{F},A^{N}\right) \right]
^{A} & =C_{BC}^{A}\mathcal{F}^{B}\varepsilon^{C}-g^{AB}C_{Bj}^{i}\mathcal{F}%
_{i}\varepsilon^{j}, \\
\left[ \mathcal{\ell}_{p+1}\left( \varepsilon,\mathcal{F},A^{p-1}\right) %
\right] ^{A} & =\left( -1\right) ^{1+\frac{\left( p-1\right) \left(
p-2\right) }{2}}C_{AB_{2}\cdots
B_{p+1}}^{i_{1}}\varepsilon^{B_{2}}A^{B_{3}}\cdots A^{B_{p+1}}\mathcal{F}%
_{i_{1}}.
\end{align}

\subsubsection{Extended sector}

We now compute the variation of $\mathcal{F}^{i}$:%
\begin{align}
\delta\mathcal{F}^{i} & =\sum_{m,n}ng^{ii_{1}}g_{A_{1}\cdots
A_{m}i_{1}\cdots i_{n}}\left(
mC_{BC}^{A_{1}}R^{B}\varepsilon^{C}R^{A_{2}}\cdots R^{A_{m}}R^{i_{1}}\cdots
R^{i_{n-1}}\right.  \notag \\
& +\left( n-1\right) R^{A_{1}}\cdots
R^{A_{m}}C_{Bj}^{i_{1}}R^{B}\varepsilon^{j}R^{i_{2}}\cdots
R^{i_{n-1}}-\left( n-1\right) R^{A_{1}}\cdots
R^{A_{m}}C_{Bj}^{i_{1}}\varepsilon^{B}R^{j}R^{i_{2}}\cdots R^{i_{n-1}} 
\notag \\
& \left. -\frac{\left( n-1\right) }{\left( p-1\right) !}R^{A_{1}}\cdots
R^{A_{m}}C_{B_{1}\cdots
B_{p+1}}^{i_{1}}\varepsilon^{B_{1}}R^{B_{2}}A^{B_{3}}\cdots
A^{B_{p+1}}R^{i_{2}}\cdots R^{i_{n-1}}\right) .   \label{dfi}
\end{align}
Using again the invariant tensor conditions (\ref{it1}) - (\ref{it3}) we
prove the following relations 
\begin{align}
0 & =g_{A_{1}\cdots A_{m}i_{1}\cdots i_{n}}\left(
mC_{BC}^{A_{1}}R^{B}\varepsilon^{C}R^{A_{2}}\cdots
R^{A_{m}}f^{i_{1}}R^{i_{2}}\cdots R^{i_{n}}-\left( n-1\right)
R^{A_{1}}\cdots
R^{A_{m}}f^{i_{1}}C_{Bj}^{i_{2}}\varepsilon^{B}R^{j}R^{i_{3}}\cdots
R^{i_{n}}\right)  \notag \\
& -g_{A_{1}\cdots A_{m}ii_{2}\cdots i_{n}}R^{A_{1}}\cdots
R^{A_{m}};C_{Bi_{1}}^{i}\varepsilon^{B}R^{i_{2}}\cdots R^{i_{n}}, 
\label{id4}
\end{align}%
\begin{equation}
0=g_{A_{1}\cdots A_{m}i_{1}\cdots i_{n}}R^{A_{1}}\cdots
R^{A_{m}}C_{B_{1}\cdots
B_{p+1}}^{i_{2}}\varepsilon^{B_{1}}R^{B_{2}}A^{B_{3}}\cdots
A^{B_{p+1}}R^{i_{3}}\cdots R^{i_{n}},   \label{id5}
\end{equation}%
\begin{equation}
0=g_{A_{1}\cdots A_{m}i_{1}\cdots i_{n}}R^{A_{1}}\cdots
R^{A_{m}}C_{Bj}^{i_{2}}R^{B}\varepsilon^{j}R^{i_{3}}\cdots R^{i_{n}}. 
\label{id6}
\end{equation}
By replacing Eqs. (\ref{id4}) - (\ref{id6}) into Eq. (\ref{dfi}) we obtain
that%
\begin{equation}
\delta\mathcal{F}^{i}=g^{ij}g_{kl}C_{Bj}^{k}\varepsilon^{B}\mathcal{F}^{l}. 
\label{deltafi}
\end{equation}
Finally, by comparing Eq. (\ref{deltafi}) with the general expression in Eq.
(\ref{df}), we obtain one consistency product for the extended sector%
\begin{equation}
\left[ \mathcal{\ell}_{2}\left( \varepsilon,\mathcal{F}\right) \right]
^{i}=g^{ij}g_{kl}C_{Bj}^{k}\varepsilon^{B}\mathcal{F}^{l}.
\end{equation}
This completes the obtention of the $L_{\infty}$ products acting on $X_{-2}$.


\begin{thebibliography}{99}
\bibitem{Zwie} O. Hohm and B. Zwiebach, \textquotedblleft$L_{\infty}$\textit{%
\ Algebras and Field Theory\textquotedblright}, Fortsch. Phys. \textbf{65}
(2017) no.3-4, 1700014, arXiv:1701.08824 [hep-th].

\bibitem{boots} R. Blumenhagen, I. Brunner, V. Kupriyanov and D. L\"{u}st,
\textquotedblleft Bootstrapping non-commutative gauge theories from L$%
_{\infty}$ algebras,\textquotedblright\ \textit{JHEP} \textbf{1805} (2018)
097, arXiv:1803.00732v2 [hep-th].

\bibitem{lada} T. Lada and J. Stasheff, \textquotedblleft Introduction to SH
Lie algebras for physicists,\textquotedblright\ \textit{Int. J. Theor. Phys.}
\textbf{32} (1993) 1087, arXiv:hep-th/9209099.

\bibitem{lada2} T. Lada and M. Markl, \textquotedblleft Strongly homotopy
Lie algebras,\textquotedblright\ \textit{Commun. Alg.} \textbf{23} (1995)
2147, arXiv:hep-th/9406095.

\bibitem{baez} J. C. Baez and J. Huerta, \textquotedblleft An invitation to
higher gauge theory,\textquotedblright\ Gen. Rel. Grav. \textbf{43} (2011)
2335, arXiv:1003.4485 [hep-th].

\bibitem{teit} C. Teitelboim, \textquotedblleft Gauge invariance for
extended objects,\textquotedblright\ \textit{Phys. Lett. B} \textbf{167}
(1986) 63.

\bibitem{dauria1} R. D'Auria, P. Fre, \textquotedblleft Geometric
Supergravity in d = 11 and Its Hidden Supergroup,\textquotedblright\ \textit{%
Nucl. Phys. B} \textbf{201} (1982) 101-140.

\bibitem{achu} A. Achucarro and P.K. Townsend, \textquotedblleft A
Chern-Simons Action for Three-Dimensional anti-De Sitter Supergravity
Theories,\textquotedblright\ \textit{Phys. Lett. B }\textbf{180} (1986) 89.

\bibitem{nie} P. Van Nieuwenhuizen, \textquotedblleft D=3 Conformal
Supergravity and Chern-simons Terms,\textquotedblright\ \textit{Phys. Rev. D}
\textbf{32} (1985) 872.

\bibitem{wi1} E. Witten, \textquotedblleft(2+1)-Dimensional Gravity as an
Exactly Soluble System,\textquotedblright\ \textit{Nucl. Phys. B} \textbf{311%
} (1988) 46.

\bibitem{cham} A. H. Chamseddine, \textquotedblleft Topological Gauge Theory
of Gravity in Five-dimensions and All Odd Dimensions,\textquotedblright\ 
\textit{Phys. Lett. B} \textbf{233} (1989) 291.

\bibitem{cham2} A. H. Chamseddine, \textquotedblleft Topological gravity and
supergravity in various dimensions,\textquotedblright\ \textit{Nucl. Phys. B}
\textbf{346} (1990) 213.

\bibitem{ban} M. Ba\~{n}ados, R. Troncoso and J. Zanelli, \textquotedblleft
Higher dimensional Chern-Simons supergravity,\textquotedblright\ \textit{%
Phys. Rev. D} \textbf{54} (1996) 2605, arXiv:gr-qc/9601003.

\bibitem{tron2} R. Troncoso and J. Zanelli, \textquotedblleft Gauge
Supergravities for All Odd Dimensions,\textquotedblright\ \textit{Int. Jour.
Theor. Phys.} \textbf{38} (1999) 1181-1206, arXiv:hep-th/9807029.

\bibitem{savv1} I. Antoniadis and G. Savvidy, \textquotedblleft New gauge
anomalies and topological invariants in various
dimensions,\textquotedblright\ \textit{Eur. Phys. J. C }\textbf{72} (2012)
2140, arXiv:1205.0027 [hep-th].

\bibitem{savv2} I. Antoniadis and G. Savvidy, \textquotedblleft Extension of
Chern-Simons forms and new gauge anomalies,\textquotedblright\ \textit{Int.
Jour. Mod. Phys. A} \textbf{29} (2014) 1450027, arXiv:1304.4398 [hep-th].

\bibitem{snpb} P. Salgado and S. Salgado, \textquotedblleft Extended gauge
theory and gauged free differential algebras,\textquotedblright\ \textit{%
Nucl. Phys. B} \textbf{926} (2018) 179, arXiv:1702.07819 [hep-th].

\bibitem{cohom} R. D'Auria, P. Fre, T. Regge, \textquotedblleft Graded Lie
Algebra Cohomology and Supergravity,\textquotedblright\ \textit{Riv. Nuovo
Cim.} \textbf{3N12} (1980) 1.

\bibitem{CSFDA} S. Salgado, \textquotedblleft Gauge-invariant theories and
higher-degree forms,\textit{\textquotedblright }\ \textit{JHEP} \textbf{09}
(2014) 055, arXiv:1310.7185 [hep-th].

\bibitem{castellaniB} L. Castellani, R. D'Auria and P. Fre,
\textquotedblleft Supergravity and superstrings: A Geometric perspective.
Vol. 2: Supergravity,\textquotedblright\ \textit{\ Singapore: World
Scientific} (1991) 607-1371.

\bibitem{kaj} H. Kajiura and J. Stasheff, \textquotedblleft Homotopy
algebras inspired by classical open-closed string field
theory,\textquotedblright\ \textit{Commun. Math. Phys.} \textbf{263} (2006)
553, arXiv:math/0410291 [math-qa].

\bibitem{cast1} L. Castellani and A. Perotto, \textquotedblleft Free
differential algebras: Their use in field theory and dual
formulation,\textquotedblright\ \textit{Lett. Math. Phys.} \textbf{38}
(1996) 321-330, arXiv:hep-th/9509031.

\bibitem{cast2} L. Castellani, \textquotedblleft Lie Derivatives along
Antisymmetric Tensors, and the M-Theory Superalgebra,\textquotedblright\ 
\textit{J. Phys. Math.} \textbf{3} (2011) P110504.

\bibitem{cast3} L. Castellani, \textquotedblleft Extended Lie derivatives
and a new formulation of D=11 supergravity,\textquotedblright\ \textit{J.
Phys. Math.} \textbf{3} (2011) P110505, arXiv:hep-th/0604213.

\bibitem{cast4} L. Castellani, \textquotedblleft Higher form gauge fields
and their nonassociative symmetry algebras,\textquotedblright\ \textit{JHEP} 
\textbf{10} (2021) 066,  arXiv:2108.02284v2 [hep-th].

\bibitem{cw} F. Izaurieta, E. Rodriguez and P. Salgado, \textquotedblleft On
transgression forms and Chern-Simons (super)gravity,\textquotedblright\
arXiv:hep-th/0512014.

\bibitem{tf} P. Mora, R. Olea, R. Troncoso and J. Zanelli, \textquotedblleft
Transgression forms and extensions of Chern-Simons gauge
theories,\textquotedblright\ \textit{JHEP} \textbf{02} (2006) 067,
arXiv:hep-th/0601081.

\bibitem{bonezzi} R. Bonezzi and O. Hohm, \textquotedblleft Leibniz gauge
theories and infinity structures,\textquotedblright\ \textit{Comm. Math.
Phys.} \textbf{377} (2020), 2027--2077, arXiv:1904.11036 [hep-th].
\end{thebibliography}
\end{document}